\documentclass[10pt,french,english,twocolumn]{IEEEtran}

\usepackage[T1]{fontenc}
\usepackage[latin9]{inputenc}
\usepackage{amsmath}
\usepackage{amssymb}
\usepackage{graphicx}
\usepackage{babel}

\makeatletter

\providecommand{\tabularnewline}{\\}

\providecommand{\tabularnewline}{\\}

\makeatother

\begin{document}

\title{Control of Multiple Remote Servers\\
for Quality-Fair Delivery of Multimedia Contents}

\author{\IEEEauthorblockN{Nesrine Changuel$^{\star}$, Bessem Sayadi$^{\star}$,
Michel Kieffer $^{o,+,\dagger}$}\\
 $^{\star}$ \IEEEauthorblockA{Alcatel-Lucent - Bell-Labs,  91620 Nozay
\\
 Email: \{nesrine.changuel, bessem.sayadi\}@alcatel-lucent.com} \\
 \IEEEauthorblockA{$^{o}$ L2S, CNRS - SUPELEC - Univ Paris-Sud, 91192 Gif-sur-Yvette \\
 $^{+}$ partly on leave at LTCI - CNRS - Telecom Paris-Tech, 75014 Paris \\
 $\dagger$ Institut Universitaire de France, 75005 Paris\\
 Email: kieffer@lss.supelec.fr}}
\maketitle
\begin{abstract}

This paper proposes a control scheme for the quality-fair delivery of
several encoded video streams to mobile users sharing a common wireless
resource. Video quality fairness, as well as similar delivery delays
are targeted among streams. The proposed controller is implemented
within some aggregator located near the bottleneck of the network.
The transmission rate among streams is adapted based on the quality
of the already encoded and buffered packets in the aggregator. Encoding
rate targets are evaluated by the aggregator and fed back to each
remote video server (fully centralized solution), or directly evaluated by each server in a distributed
way (partially distributed solution). Each encoding rate target is adjusted for each stream independently
based on the corresponding buffer level or buffering delay in the
aggregator. Communication delays between the servers and the aggregator
are taken into account.

The transmission and encoding rate control problems are studied with
a control-theoretic perspective. The system is described with a multi-input
multi-output model. Proportional Integral (PI) controllers are used
to adjust the video quality and control the aggregator buffer levels.
The system equilibrium and stability properties are studied. This
provides guidelines for choosing the parameters of the PI controllers.

Experimental results show the convergence of the proposed control
system and demonstrate the improvement in video quality fairness compared
to a classical transmission rate fair streaming solution and to a utility
max-min fair approach\footnote{Parts of this work have been presented at ACM Multimedia conference, 2012. This work has been partly supported by ANR ARSSO project, contract number ANR-09-VERS-019-02 and by ANR project LimICoS, contract number ANR-12-BS03-005-01.}.
\end{abstract}


\section{Introduction}

\label{sec:intro}

With the development of wireless networks and widespread of smartphones,
delivery of compressed videos (video-on-demand or mobile TV broadcast
services) to mobile users is increasing rapidly. This trend is likely
to continue in the coming years~\cite{Cisco2009}. To satisfy the
related increasing demand for resources, operators have to expand
their network capacity with as limited as possible infrastructure
investments. In parallel, they have to optimize the way multimedia
contents are delivered to users while satisfying \emph{application-layer}
quality-of-service (QoS) constraints, which are more challenging to
address than traditional \emph{network-layer} QoS constraints.

Delivered videos have a large variety of quality-rate characteristics,
whatever the considered quality metric, \emph{e.g.}, Peak Signal-to-Noise
Ratio (PSNR), Structural SIMilarity (SSIM)~\cite{Wang2004}, \emph{etc}.
These characteristics are time-varying and depend on the content of
the videos. Provisioning some constant transmission rate to mobile
users for video delivery is in general inappropriate. If videos are
encoded at a constant bitrate, the quality may fluctuate with the
variations of the characteristics of the content. If they are encoded
at a variable bitrate, targeting a constant quality, buffering delays
may fluctuate significantly.

This paper proposes a controller for the quality-fair delivery of
several encoded video streams to mobile users sharing a common wireless
resource. Video-on-demand or multicast/broadcast transmission are typical
applications for this scenario. Video encoding rate adaptation and
wireless resource allocation are performed jointly within some Media
Aware Network Element (MANE) using feedback control loops. The aim
is to provide users with encoded videos of similar quality and with
controlled delivery delay, without exchanging information between
remotely located video servers.

\subsection{Related work}

When controlling the parallel delivery of several video streams, their
rate-distortion (R-D) trade-off may be adjusted by selectively discarding
frames as in~\cite{Zhang2001,Lin2009} or via an adaptation of their
encoding parameters as in~\cite{Ma2005,He2008}. With scalable video
encoders, such as H.264/SVC, the R-D trade-off may be adjusted via
packet filtering~\cite{Srinivasan2010,Maani2010}. In this case, the
control parameter is the number of transmitted enhancement layers for each frame.

If several video streams are transmitted to different users in a dedicated
broadcast channel with limited capacity, a blind source rate allocation
could lead to unacceptable quality for high-complexity video contents
compared to low-complexity ones. Therefore, providing fairness is
an important issue that must be addressed.

Video quality fairness among encoded streams may be obtained by sharing
quality information, or R-D characteristics via a central controller
providing to each server quality or rate targets, as in~\cite{Li2009a,Changuel2010a}.
This technique enables the encoders to adjust their bit rate or to
drop frames or quality layers depending on the complexity of the videos and on the available
transmission rate.

Control-theoretic approaches have been considered to address the problem
of rate control in the context of video streaming, see, \emph{e.g.}, \cite{Wong2004,Guo2005,Zhang2008,Huang2009a,Huang2009}.
In \cite{Wong2004}, a real-time rate control based on a Proportional
Integral Derivative (PID) controller is proposed for a single video
stream. The main idea is to determine the encoding rate per frame
based on the buffer level to maximize video quality and minimize quality
variations over frames. In~\cite{Guo2005}, a flow control mechanism
with active queue management and a proportional controller is considered.
A flow control is used to reduce the buffer size and avoid buffer
overflow and underflow. The flow control mechanism is shown to be
stable for small buffer sizes and non-negligible round-trip times.
In~\cite{Zhang2008}, a rate allocation algorithm, performed at the
Group of Picture (GoP) level, is performed at the sender to maximize
the visual quality according to the overall loss and the receiver
buffer occupancy. This target is achieved using a Proportional Integral
(PI) controller in charge of determining the transmission rate to
drain the buffers. Later,~\cite{Zhou2011} introduces a rate controller
that uses different bit allocation strategies for Intra and Inter
frames. A PID controller is adopted to minimize the deviations between
the target and the current buffer level. Buffer management is performed
at the bit level and delivery delay is not considered. Moreover,
\cite{Wong2004,Guo2005,Zhang2008,Zhou2011} address single flow transmission.

For multi-video streaming, in~\cite{Cho2005}, a distributed utility
max-min flow control in the presence of round-trip delays is proposed.
The distributed link algorithm performs utility max-min bandwidth sharing
while controlling the link buffer occupancy around a target value at the cost
of link under utilization using a PID controller. Stability analysis
in case of a single bottleneck and homogeneous delay is conducted.
In~\cite{Li2009a}, a content-aware distortion-fair video delivery
scheme is proposed to deliver video streams based on the characteristics
of video frames. It provides a max-min distortion fair resource sharing
among video streams. The system uses temporal prediction structure
of the video sequences with a frame drop strategy based on the frame
importance to guide resource allocation. The proposed scheme is for
video on-demand services, where the rate and the importance of each
frame are assumed calculated in advance. A proportional controller
is considered in~\cite{Huang2009a} to stabilize the received video
quality as well as the bottleneck link queue for both homogeneous
and heterogeneous video contents. A PI controller is considered in~\cite{Huang2009}.
Robustness and stability properties are studied. In~\cite{Huang2009a}
and~\cite{Huang2009}, the rate control is performed in a centralized
way, exploiting the rate and distortion characteristics of the considered
video flows to determine the encoding rate for the next frame of each
flow. In~\cite{Cicalo2012}, a cross-layer optimization framework
for scalable video delivery over OFDMA wireless systems is proposed,
aiming at maximizing the sum of the achievable rates while minimizing
the distortion difference among multiple videos. The optimization
problem is described by a Lagrangian constrained sum-rate maximization
to achieve distortion fairness among users. However the communication
delay between the control block and the controlled servers is not
addressed.

Remotely implemented control laws are also considered
in~\cite{Witrant2007} leading to the problem of stabilizing an open-loop
system with time-varying delay. The problem of remote stabilization
via communication networks is considered with an explicit use of the
average network dynamics and an estimation of the average delay in
the control law. The control law does not address video transmission
issues, so no quality constraint on the transmitted data is considered.

The commercial products described in~\cite{Motorola2009,Harmonic2011}
propose statistical multiplexing systems enabling encoders to adapt
their outputs to the available channel rate. Connecting encoders and
multiplexers via a switched IP network allows collocated and distributed
encoders to be part of the multiplexing system. Nevertheless, in these
solutions, quality fairness constraints between programs appear not
to be considered among the video quality constraints.

\subsection{Main contributions}

In this paper, we propose a control system to perform jointly (\emph{i})
encoding rate control of spatially spread video servers without information
exchange between them and (\emph{ii}) transmission rate control of
the encoded streams through some bottleneck link. A MANE, located
near the bottleneck link derives the average video quality of the
data stored in dedicated buffers fed by the remote servers. This
average video quality is compared by each individual transmission
rate controller to the quality of its video flow to adjust the draining
rate of the corresponding buffer (first control input). For that purpose,
programs with low quality are drained faster than programs with high
quality. Dedicated encoding rate controllers observe the buffer levels
to adjust the video encoding rates (second control input). The encoding
rate control targets a similar buffer level for all programs. The
buffer level in bits or the buffering delay can be adjusted via an adaptation
of the video encoding rates, \emph{e.g.}, by scalability layer filtering when a
scalable video coder is involved.

In a \emph{fully centralized} version of the controller, the MANE
is in charge of sending the encoding rate target to each video server.
In a \emph{partly distributed} version, only the individual buffer
level discrepancies are transmitted to the servers, which are then
in charge of computing their own encoding rate target. Communication
delays between the MANE and the servers are considered in both directions.
A discrete-time state-space representation of the system is introduced.
The buffer level (in bits) or the buffering delay has to be controlled
and quality fairness between video streams has to be obtained. For
that purpose, feedback loops involving PI controllers are considered.
The quality fairness constraint among streams leads to a coupling
of the state equations related to the control of the delivery of each
stream. The system equilibrium and stability properties are studied.
This provides guidelines for choosing the parameters of the PI controllers.
This paper extends preliminary results obtained in \cite{Changuel2012a},
where control of the buffer level in bits is not addressed and the
communication delay between the MANE and the servers is not explicitly
considered.

Section~\ref{sec:problem} introduces the considered system and the
constraints that have to be satisfied. The proposed solution is described
in Section~\ref{sec:ProposedSolution}. The required hypotheses are
listed and a discrete-time state-space representation of the system
is provided, emphasizing the coupling between equations induced by
the fairness constraint among video streams. The equilibrium and stability
analyses are performed in Section~\ref{sec:EqandStab}. Finally,
a typical application context is described in Section~\ref{sec:exp}
and experimental results are detailed. Robustness of the proposed
control system to variations of the channel rate and of the number
of video streams is shown.

\section{System description}
\label{sec:problem}

\begin{figure}[t!pb]
 \centering \includegraphics[width=\columnwidth]{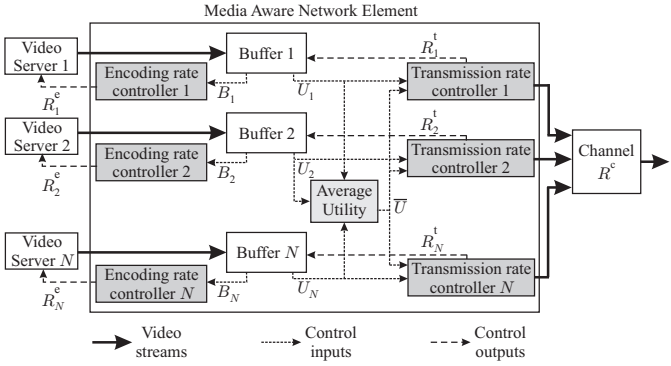}
\caption{Structure of the proposed quality-fair video delivery system (fully
centralized version).}
\label{fig:diagram}
\end{figure}

Consider a communication system in which $N$ encoded video streams
provided by $N$ remote servers arrive to some network bottleneck
where they have to share a communication channel providing a total
transmission rate $R^{\text{c}}$, see Figure~\ref{fig:diagram}.
The servers deliver encoded Video Units (VUs) representing a single
picture or a Group of Pictures (GoP). All VUs are assumed to be of
the same duration $T$ and the frame rate $F$ is assumed constant
over time and identical for all streams. Time is slotted and the $j$-th
time index represents the time interval $\left[jT,\left(j+1\right)T\right)$.

\begin{figure}[htpb]
\centering \includegraphics[width=0.9\columnwidth]{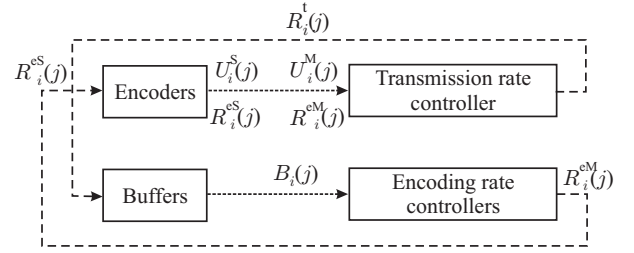}
\caption{Feedback control loops in the proposed quality-fair video delivery
system.}
\label{fig:control1}
\end{figure}

In the proposed system, a Media Aware Network Element (MANE) is located
close to the bottleneck of the network, see Figure~\ref{fig:diagram}.
The MANE aims at providing the receivers with video streams of similar
(objective or subjective) quality and with similar delivery delays.
For that purpose, two feedback loops are considered to control \emph{(i)
}the encoding rate and \emph{(ii)} the transmission rate of each video
stream, see Figure~\ref{fig:control1}.

Encoded and packetized VUs are temporarily stored in dedicated buffers
in the MANE. We assume that a \emph{utility} $U_{i}\left(j\right)$ measures
the quality of each VU $j$ for each stream
$i$ (in terms of PSNR, SSIM, or any other video quality metric \cite{Seshadrinathan2010}).
The MANE has access to $U_{i}\left(j\right)$, stored, \emph{e.g.},
in the packet headers. The transmission rate controllers are in charge
of choosing the draining rates from each buffer so that all utilities
within the $N$ buffers are as close as possible. The encoding rate
controllers are in charge of choosing the video encoding rates so that
the buffer levels in the MANE are adjusted around some reference level
$B_{0}$ in bits or reference delay $\tau_{0}$ in seconds.
This control is performed at each discrete time index.

The MANE evaluates the average utility and each transmission rate controller allocates more
rate to streams with a utility below average. The buffers of these
streams are drained faster than those with utility above average.
This control, done within the MANE, is thus \emph{centralized}. The
discrepancy of each buffer level in bits (respectively delay in seconds)
with respect to the reference level $B_{0}$ (respectively delay $\tau_{0}$)
is processed individually by an encoding rate controller. The encoding
rate for the next VU of each video stream is then evaluated to regulate
the buffer level around the reference level. The encoding rate may
be evaluated at the MANE, in which case, the control is fully \emph{centralized}
and the video encoders/servers receive only the evaluated
target bit rate. Alternatively, the encoding rates may be evaluated
in a \emph{decentralized} way at each video encoder/server. For the
latter case, the MANE only feeds back to each server the buffer level
(respectively delay) discrepancy. This solution requires all video
encoders/servers to host individually an encoding rate controller,
which is mainly possible for managed video servers. In this paper,
the encoding rate target is evaluated within the MANE, but the encoding
parameters are evaluated in a \emph{distributed} way by each server
\cite{Wiegand2003b}.

The interaction of both control loops (transmission rate control and
encoding rate control) allows getting a quality-fair video delivery.
Videos with a quality below average have a buffer in the MANE that
is drained faster, and thus is likely to be below $B_{0}$ or $\tau_{0}$.
The encoding rate of such streams is then increased, to improve their
quality.

Feedback delays between the MANE and the video servers are considered.
They correspond to the delays introduced when the servers deliver
encoded packets to the MANE and when the MANE feeds back signalization
to carry encoding rate targets to the servers. To account for the
delayed utility available at the MANE and the delayed encoding rate
targets sent by the MANE to the servers, two state variables representing
the delayed encoding rates and the delayed utilities are introduced
\begin{equation}
R_{i}^{\text{eS}}\left(j\right)=R_{i}^{\text{eM}}\left(j-\delta_{1}\right),\label{eq:DelayedEncRate}
\end{equation}
and
\begin{equation}
U_{i}^{\text{M}}\left(j\right)=U_{i}^{\text{S}}\left(j-\delta_{2}\right),\label{eq:Delayedutility}
\end{equation}
where $R_{i}^{\text{eM}}(j)$ is the encoding rate evaluated at the
MANE for the $j$-th VU of program $i$. This encoding rate target
reaches the server with some delay $\delta_{1}$, where it is denoted
by $R_{i}^{\text{eS}}(j)$. On the other hand, the server sends the
utility of the $j$-th VU of program $i$, denoted by $U_{i}^{\text{S}}(j)$.
It arrives at the MANE with some delay $\delta_{2}$ and is denoted
$U_{i}^{\text{M}}(j)$, see Figure~\ref{fig:control1}. The $\mbox{M}$
superscript refers to the information available at the MANE and the
$\mbox{S}$ superscript refers to the information available at the
server for both rate and utility. In what follows, the stability of
both control loops is studied.

\section{State-space representation}

\label{sec:ProposedSolution}

A state-space representation of the system presented in Section~\ref{sec:problem}
is introduced to study its stability. Several additional assumptions
are needed to get a tractable representation.

\subsection{Assumptions}

\subsubsection{Feedback delay}

\begin{figure}[htpb]
\centering \includegraphics[width=0.6\columnwidth]{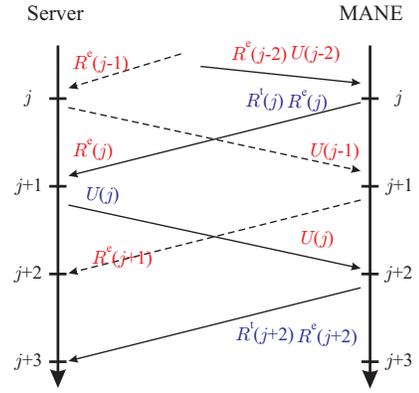}
\caption{Communication delays between a video server and the MANE.}
\label{fig:delay}
\end{figure}

In what follows, a VU represents a GoP. All encoded VUs processed
by the video servers during the $\left(j-1\right)$-th time slot $\left[(j-1)T,jT\right)$
are assumed to have reached the MANE during the $j$-th time slot
$\left[jT,\left(j+1\right)T\right)$. The encoding rate targets sent
by the MANE to the servers $R_{i}^{\text{e}}(j)$ at the beginning
of the $j$-th time slot are assumed to have reached all video servers
at the beginning of the $\left(j+1\right)$-th time slot. This rate
is used to encode the $\left(j+1\right)$-th VU which will be placed
in the buffer in the MANE during the $\left(j+2\right)$-th time slot,
\emph{etc.}, see Figure~\ref{fig:delay} when $\delta_{1}=\delta_{2}=1$.

The transmission delays between the MANE and the servers may vary.
The previous assumptions allow to cope with forward and backward delays
upper bounded by $T$. This is reasonable, since the network transmission
and buffering delays are of the order of tens of milliseconds, provided
that it is not too congested. This is less than the duration of VUs
when they represent GoPs (typically $0.25$~s to $1$~s). Following
the bounded-delay assumption, during the $j$-th time slot, the MANE
has only access to the utilities $U_{i}^{\text{S}}(j-2)$, $i=1,\dots,N$
of the $\left(j-2\right)$-th encoded VUs.

In the rest of the paper the superscripts S and M are omitted. Then
$U_{i}\left(j\right)$ is the utility of the $j$-th VU of the $i$-th
stream encoded during time slot $j$, transmitted to the MANE during
time slot $j+1$, and fully buffered in the MANE at the beginning
of time slot $j+2$. $R^{\text{e}}(j)$ is the encoding rate target
evaluated by the MANE during the $j$-th time slot. $R^{\text{e}}(j)$ reaches the
server at the beginning of time slot $j+1$, see Figure~\ref{fig:delay}.

\subsubsection{Source model}

The following parametric rate-utility model is used to describe the
evolution of the utility $U_{i}(j)$ as a function of the rate $R_{i}^{\text{e}}(j)$
used to encode the $j$-th VU of the $i$-th stream
\begin{equation}
U_{i}(j)=f\left(\mathbf{a}_{i}\left(j\right),R_{i}^{\text{e}}(j)\right),\label{Eq:psnr2}
\end{equation}
where $\mathbf{a}_{i}\left(j\right)\in\mathcal{A}\subset\mathbb{R}^{N_{a}}$
is a time-varying and program-dependent parameter vector.
Note that this model is only used to define the controller parameters so that
the system is stable. Once these parameters are set, the rate-utility model is
no more needed. For all values of $\mathbf{a}$ belonging to the set of admissible parameter
values $\mathcal{A}$, $f\left(\mathbf{a},R\right)$ is assumed to
be a continuous and strictly increasing function of $R$, with $f\left(\mathbf{a},0\right)=0$.
The variation with time of $\mathbf{a}_{i}\left(j\right)$ is described by
\begin{align}
\mathbf{a}_{i}(j+1) & =\mathbf{a}_{i}(j)+\delta\mathbf{a}_{i}(j),\label{eq:NoiseA}
\end{align}
where $\delta\mathbf{a}_{i}\left(j\right)$ represents the uncontrolled
variations of the source characteristics.

The model~\eqref{Eq:psnr2} may represent the variation with the
encoding rate of the SNR, the PSNR, the SSIM, or any other strictly
increasing quality metric.

\subsubsection{Buffer model}

\label{Ssec:BufferModel}

As introduced in Section~\ref{sec:problem}, the MANE contains dedicated
buffers for each of the $N$ encoded video streams. The evolution
of the level in bits of the $i$-th buffer between time slot $j$
and $j+1$ is
\begin{equation}
B_{i}(j+1) = B_{i}(j)+\left(R_{i}^{\text{e}}(j-2)-R_{i}^{\text{t}}(j)\right)T,\label{eq:buffer}
\end{equation}
where $R_{i}^{\text{t}}(j)$ is the transmission rate for the $i$-th
stream and $R_{i}^{\text{e}}(j-2)$ is the encoding rate of the $\left(j-2\right)$-th
VU evaluated at the MANE at time slot $j$. In~\eqref{eq:buffer},
$R_{i}^{\text{e}}(j-2)$ accounts for the communication delay between
the server and the MANE, see Figure \ref{fig:delay}.

Buffers are controlled in two different ways. A control of the
buffer level in bits $B_{i}(j)$ maintains an averaged level in bits to prevent buffer overflow and underflow. This is appropriate for applications with buffers of limited size. A control of the buffering
delay helps adjusting the end-to-end delivery delay. This type of
control is better suited for delay-sensitive applications.
Buffering delay control within the MANE also allows implicitly controlling the buffering delay at the client for live and broadcast applications. This is due to the fact that the end-to-end delay between the transmission of a VU by the server and its playback by the client is constant over time for live video\footnote{Some periodic feedback may nevertheless be useful to verify that the system actually behaves nominally.}.
% This delay consists of the delay at the transmitter, the delivery delay, and the delay at the receiver. Thus a regulated transmission delay results in a regulated client buffering delay for limited transmission delay variations.

Let $h_{i}(j)$ be the number of VUs in the $i$-th MANE buffer at
time $j$, the corresponding buffering delay is
\begin{equation}
\tau_{i}(j)=h_{i}(j)T.\label{eq:DelayEstimate1}
\end{equation}
Assume that packets containing encoded VUs are segmented to allow
a fine granularity of the draining rate. Then $h_{i}(j)$ becomes
quite difficult to evaluate accurately and directly within the MANE using only information
stored in packet headers. The corresponding buffering delay is then
approximatively evaluated as
\begin{equation}
\tau_{i}(j)=\frac{B_{i}(j)}{\bar{R}_{i}^{\text{e}}(j)},\label{eq:DelayEstimate}
\end{equation}
where
\begin{equation}
\begin{array}{ll}
\bar{R}_{i}^{\text{e}}(j)& = \frac{1}{h_{i}(j)}\sum_{\ell=2}^{\lfloor h_{i}(j)\rfloor}R_{i}^{\text{e}}(j-\ell)\\
 &+ \frac{1}{h_{i}(j)}R_{i}^{\text{e}}\left(j-\lceil h_{i}(j)\rceil\right)\left(h_{i}(j)-\lfloor h_{i}(j)\rfloor\right)
\end{array} \label{eq:AvgRate}
\end{equation}
is the average encoding rate of the VUs stored in the $i$-th buffer
at time $j$ and $\left\lfloor \cdot\right\rfloor $ and $\left\lceil \cdot\right\rceil $
denote rounding towards $-\infty$ and $+\infty$. Since~\eqref{eq:AvgRate}
still requires the availability of $h_{i}(j)$, we propose to estimate
it as follows
\begin{equation}
\begin{array}{l}
\tilde{R_{i}}^{\text{e}}(j)=R_{i}^{\text{e}}(j),\text{ if }j \leqslant 2,\\
\tilde{R_{i}}^{\text{e}}(j)=\alpha R_{i}^{\text{e}}(j-2)+(1-\alpha)\tilde{R_{i}}^{\text{e}}(j-1),\text{ if } j > 2,
\end{array}\label{eq:AvgRateEstim}
\end{equation}
where $0<\alpha<1$ is some tuning parameter. Then, one gets an estimate
of the buffering delay~\eqref{eq:DelayEstimate1} using~\eqref{eq:AvgRateEstim}
\begin{equation}
\tilde{\tau}_{i}(j)=\frac{B_{i}(j)}{\tilde{R}_{i}^{\text{e}}(j)}.\label{eq:BufferingDelayEstim}
\end{equation}

\subsection{Rate controllers}

$N$ coordinated transmission rate controllers and $N$ (possibly distributed)
encoding rate controllers are considered in the video delivery system
described in Section~\ref{sec:problem}. A PI control of the transmission
rate is performed. At time $j$, each PI controller takes as input
the average utility evaluated by the MANE and the utility $U_{i}\left(j-2\right)$ of the VU $j-2$ of the controlled stream.
PI controllers are also used to evaluate the target encoding rate of
each program in order to regulate the buffer level around $B_{0}$
or the buffering delay around $\tau_{0}$.

\subsubsection{Transmission rate control}

At time $j$, the available channel rate $R^{\text{c}}$ is shared
between the $N$ video streams. The delayed utility of the VU available
at the MANE at time $j$ for the $i$-th stream is
\begin{equation}
U_{i}^{\text{dd}}(j)=U_{i}^{\text{d}}(j-1)=U_{i}(j-2)\label{eq:Delayedutility2}
\end{equation}
 and the utility discrepancy $\delta U_{i}^{\text{dd}}(j)$ with the
\emph{average} utility $\bar{U}(j)$ over the $N$ programs at time $j$ is
\begin{equation}
\delta U_{i}^{\text{dd}}(j)=\frac{1}{N}\sum_{\ell=1}^{N}\left(U_{\ell}^{\text{dd}}(j)-U_{i}^{\text{dd}}(j)\right)=\bar{U}(j)-U^{dd}_i(j).\label{eq:PSNR discrepancy}
\end{equation}

The PI transmission rate controller for the $i$-th program uses $\delta U_{i}^{\text{dd}}(j)$
to evaluate the transmission rate allocated to each video stream
\begin{equation}
R_{i}^{\text{t}}\left(j\right)=R_{0}+\left(K_{P}^{\text{t}}+K_{I}^{\text{t}}\right)\delta U_{i}^{\text{dd}}(j)+K_{I}^{\text{t}}\phi_{i}\left(j\right),\label{eq:transmission}
\end{equation}
where $K_{\text{P}}^{\text{t}}$ and $K_{\text{I}}^{\text{t}}$ are
the proportional and integral correction gains. All rates are evaluated
with respect to $R_{0}=R^{\text{c}}/N$, the average encoding rate
per stream that would be used when the $N$ streams represent the
same encoded video. In~\eqref{eq:transmission}, $\phi_{i}\left(j\right)$
is the cumulated utility discrepancy (used for the integral term)
evaluated as
\begin{equation}
\begin{array}{l}
\phi_{i}\left(j\right) =0,\text{ if }j \leqslant 2,\\
\phi_{i}\left(j+1\right) =\phi_{i}\left(j\right)+\delta U_{i}^{\text{dd}}(j),\text{ if } j > 2.
\end{array}
\label{eq:phi}
\end{equation}
 $R_{i}^{\text{t}}\left(j\right)$ represents the draining rate of
the $i$-th MANE buffer at time $j$. One may easily verify that
\begin{equation}
\sum_{i=1}^{N}R_{i}^{\text{t}}\left(j\right)=R^{\text{c}}.\label{eq:RateConstraint}
\end{equation}

According to~\eqref{eq:transmission}, in a first approximation, more
(resp. less) transmission rate is allocated to programs with a utility
below (resp. above) average.

\subsubsection{Encoding rate control}

\label{Ssec:EncRateControl}

The encoding rate control is performed independently for each video
stream. This allows a distributed implementation of this part of the
global controller.

For the control of the buffer level or of the buffering delay, the
buffer level~\eqref{eq:buffer} is needed. Let $R_{i}^{\text{edd}}(j)$
be the delayed encoding rate of the VU reaching the MANE at time $j$
for the $i$-th program
\begin{equation}
R_{i}^{\text{edd}}(j)=R_{i}^{\text{ed}}(j-1)=R_{i}^{\text{e}}(j-2).\label{eq:DelayedEncRate2}
\end{equation}
 Using~\eqref{eq:buffer} and~\eqref{eq:DelayedEncRate2}, one gets
\begin{equation}
\begin{array}{ll}
B_{i}(j+1) & =B_{i}(j)+\left(R_{i}^{\text{edd}}(j)-R_{i}^{\text{t}}(j)\right)T.\end{array}\label{eq:BufferDealyedEncRate}
\end{equation}

In case of buffer level control, the encoding rate for the $j$-th
VU of each video program is controlled to limit the deviations of
$B_{i}(j)$ from the reference level $B_{0}$. At time $j$, the discrepancy
$\delta B_{i}\left(j\right)$ between $B_{i}\left(j\right)$ and $B_{0}$
is
\begin{equation}
\delta B_{i}\left(j\right)=B_{i}\left(j\right)-B_{0}.
\end{equation}

In case of buffering delay control, the encoding rate for the $j$-th
VU of each video program is controlled to limit the deviations of
$\tilde{\tau}_{i}\left(j\right)$ from the reference level $\tau_{0}$.
At time $j$, the discrepancy $\delta \tau_{i}\left(j\right)$ between
$\tilde{\tau}_{i}\left(j\right)$ and $\tau_{0}$ is
\begin{equation}
\begin{array}{l}
\delta \tau_{i}\left(j\right)=\tilde{\tau}_{i}\left(j\right)-\tau_{0}=\left(\frac{B_{i}\left(j\right)}{\tilde{R}_{i}^{\text{edd}}(j)}-\tau_{0}\right).\end{array}\label{eq:tau-1}
\end{equation}
Buffers with positive $\delta B_{i}\left(j\right)$ or $\delta \tau_{i}\left(j\right)$ contain more than
$B_{0}$ bits or $\tau_{0}$ seconds of encoded videos. The encoding rate $R_{i}^{\text{e}}\left(j\right)$
has thus to be decreased. This part of the control process is very
similar to back-pressure algorithms \cite{Kim1996}.

$R_{i}^{\text{e}}\left(j\right)$ is evaluated as the output of a
PI controller
\begin{equation}
R_{i}^{\text{e}}\left(j\right)=R_{0}-\frac{K_{P}^{\text{e}}+K_{I}^{\text{e}}}{T}\delta \tau_{i}\left(j\right)-\frac{K_{I}^{\text{e}}}{T}\Pi_{i}\left(j\right).\label{eq:encodingb}
\end{equation}
where $K_{\text{P}}^{\text{e}}$ and $K_{\text{I}}^{\text{e}}$ are
the proportional and integral gains and $\Pi_{i}\left(j\right)$ is
the cumulated buffer discrepancy in seconds evaluated as
\begin{equation}
\begin{array}{l}
\Pi_{i}\left(j\right) =0,\text{ if } j\leqslant 3,\\
\Pi_{i}\left(j+1\right) =\Pi_{i}\left(j\right)+\delta \tau_{i}\left(j\right),\text{ if } j > 3.
\end{array}\label{eq:Pib}
\end{equation}
For a control of the buffer level, $\delta \tau_{i}\left(j\right)$ is replaced by $\delta B_{i}\left(j\right)$ in \eqref{eq:encodingb} and \eqref{eq:Pib}.

Taking into account the communication delay between the MANE and
the server, the encoding rate target $R_{i}^{\text{e}}\left(j\right)$ evaluated
at the MANE at time $j$ reaches the video server at time $j+1$.
Thus, $R_{i}^{\text{e}}\left(j\right)$ represents the encoding rate
for the $j+1$-th VU. The encoding rate increases (resp. decreases)
when the buffer is below (resp. above) its reference level. The sum of the encoding rates is not necessarily equal
to $R^{\text{c}}$. This allows to compensate for the variations of
the video characteristics.

Considering simultaneously \eqref{eq:buffer}, \eqref{eq:transmission},
and \eqref{eq:encodingb}, one sees that buffers corresponding to
programs producing video with lower utility than average are drained
faster. As a consequence, the encoding rate allowed to encode the
next VU of such programs is increased, potentially increasing the
utility.

\subsection{State-space representation}

The state-space representation facilitates the study of the system
equilibrium and stability properties. Two representations are considered,
depending on whether the buffer level or the buffering delay is controlled.

In the case of a control of the buffering delay, combining~\eqref{Eq:psnr2},
\eqref{eq:NoiseA}, \eqref{eq:Delayedutility2}, \eqref{eq:transmission},
\eqref{eq:phi}, \eqref{eq:BufferDealyedEncRate}, \eqref{eq:encodingb},
and \eqref{eq:Pib} leads to the following discrete-time nonlinear state-space
representation for the $i$-th video stream, $i=1,\dots,N$
%\scalebox{0.8}{$
%\left\{
\begin{subequations}\label{eq:stateSpace1}
 \begin{align}
\boldsymbol{a}_{i}(j+1)  & =\boldsymbol{a}_{i}(j)+\delta\boldsymbol{a}_{i}(j) \label{Paremeter}\\
 \boldsymbol{a}_{i}^{\text{d}}(j+1) & =\boldsymbol{a}_{i}(j)\label{Pvar}\\
 \phi_{i}\left(j+1\right) &  =\phi_{i}\left(j\right)+\frac{1}{N}\sum_{k=1}^{N}U_{k}^{\text{dd}}\left(j\right)-U_{i}^{\text{dd}}\left(j\right)\label{Phi}\\
 \Pi_{i}^{\tau}\left(j+1\right) &  =\Pi_{i}^{\tau}\left(j\right)+\left(\frac{B_{i}\left(j\right)}{\tilde{R}_{i}^{\text{e}}(j)}-\tau_{0}\right)\label{Pi}\\
 \tilde{R_{i}}^{\text{e}}(j+1) &  =\alpha R_{i}^{\text{edd}}(j)+(1-\alpha)\tilde{R_{i}}^{\text{e}}(j)\label{Retilde}\\
 R_{i}^{\text{ed}}(j+1) &  =R_{0}-\frac{K_{P}^{\text{e}\tau}+K_{I}^{\text{e}\tau}}{T}\left(\frac{B_{i}\left(j\right)}{\tilde{R}_{i}^{\text{e}}(j)}-\tau_{0}\right)-\frac{K_{I}^{e}\tau}{T}\Pi_{i}\left(j\right)\label{Red}\\
 R_{i}^{\text{edd}}(j+1)  & =R_{i}^{\text{ed}}(j)\label{redd}\\
 U_{i}^{\text{dd}}(j+1) &  =f\left(\boldsymbol{a}_{i}^{\text{d}}\left(j\right),R_{i}^{\text{ed}}(j)\right) \label{Udd}\\
 B_{i}(j+1) &=B_{i}(j)+R_{i}^{\text{edd}}(j)T-R_{0}T \nonumber\\
&\hspace{-1.5cm}-\left(\left(K_{P}^{\text{t}}+K_{I}^{\text{t}}\right)(\frac{1}{N}\sum_{k=1}^{N}U_{k}^{\text{dd}}\left(j\right)-U^{\text{dd}}\left(j\right))+K_{I}^{\text{t}}\phi_{i}(j)\right)T \label{Buff}
\end{align}
\end{subequations}
%\right
%$}
where $\boldsymbol{a}_{i}^{\text{d}}(j)$ is the delayed video characteristic
vector of the $(j-1)$-th VU. The utilities
$U_{k}^{\text{dd}}\left(j\right)$ of all video streams appear in~\eqref{eq:stateSpace1}, leading
to a coupling of the state-space representations related to the control
of the individual video streams.

When buffer level control is addressed, $B_{i}\left(j\right)/\tilde{R}_{i}^{\text{e}}(j)$ is replaced by $B_{i}\left(j\right)$  in~\eqref{Pi} and~\eqref{Red}, and the state~\eqref{Retilde} does
not appear anymore.

In the remainder of the paper, the subscript $b$ is for buffer level control and the subscript $\tau$ is for buffering delay control.

\section{Equilibrium and stability}

\label{sec:EqandStab}

The steady-state behavior and the stability of the video delivery
system described by~\eqref{eq:stateSpace1} for buffering delay control
as well as the simpler system for buffer level control are studied.
Due to the coupling between controllers induced by the constraint
that the discrepancy between the average utility and the utility of
each program has to be as small as possible, both characterizations
have to be done on the whole system. In the rest of this section,
we derive the equilibrium and perform the stability analysis for the control system
where buffering delays are controlled. The technique is similar when for the
control of the buffer level.

\subsection{Equilibrium analysis}
\label{Ssec:EquilBufferLevel}

The system reaches an equilibrium when all
terms on the left of the state-space representation~\eqref{eq:stateSpace1} do not change with time. This
leads to a system of $\left(N_{a}+6\right)\times N$ equations with
$\left(N_{a}+6\right)\times N$ unknowns.
\begin{equation}
\left\{ \begin{array}{lll}
\delta\boldsymbol{a}_{i}^{\text{eq}} & = & 0\\
U_{i}^{\text{eq}} & = & \frac{1}{N}\sum_{k=1}^{N}U_{k}^{\text{eq}}\\
B_{i}^{\text{eq}} & = & \tau_{0}\tilde{R_{i}}^{\text{e,eq}}\\
\tilde{R_{i}}^{\text{e,eq}} & = & R_{i}^{\text{e,eq}}\\
R_{i}^{\text{e},\text{eq}} & = & R_{0}-\frac{K_{\text{I}}^{\text{e}\tau}}{T}\Pi_{i}^{\tau\text{eq}}\\
U_{i}^{\text{eq}} & = & f\left(\boldsymbol{a}_{i}^{\text{eq}},R_{i}^{\text{e,eq}}\right)\\
R_{i}^{\text{e,eq}} & = & K_{\text{I}}^{\text{t}}\phi_{i}^{\text{eq}}
\end{array},\right.\label{eq:Equilibre2}
\end{equation}
The second equation in~\eqref{eq:Equilibre2} imposes that $U_{1}^{\text{eq}}=\dots=U_{N}^{\text{eq}}=U^{\text{eq}}$;
all programs have thus the same utility at equilibrium. Moreover,
one has $\tilde{R_{i}}^{\text{e},\text{eq}}=R_{i}^{\text{e,eq}}$
and $B_{i}^{\text{eq}}=\tau_{0}R_{i}^{\text{e,eq}}$, which leads
to $B_{i}^{\text{eq}}/R_{i}^{\text{e,eq}}=\tau_{0}$, for $i=1,\dots,N$
meaning that at equilibrium, the buffering delay is equal to $\tau_{0}$
for all streams.

The target encoding rates at equilibrium $R_{i}^{\text{e,eq}}$ and
the utility $U^{\text{eq}}$ are obtained as the solution of a system
of $N+1$ equations
\begin{equation}
\left\{\begin{array}{l}
f\left(\boldsymbol{a}_{1}^{\text{eq}},R_{1}^{\text{e,eq}}\right)=\dots=f\left(\boldsymbol{a}_{N}^{\text{eq}},R_{N}^{\text{e,eq}}\right)=U^{\text{eq}}\\
\sum_{i=1}^{N}R_{i}^{\text{e,eq}}=R^{\text{c}}
\end{array}\right.,\label{eq:resolution-1}
\end{equation}
depending of the values $\boldsymbol{a}_{i}^{\text{eq}}$,$i=1,\dots,N$
of the parameter vector of the rate-utility model. At equilibrium,
they are assumed constant in time and well-estimated, see Section~\ref{sec:exp}.
Since $f$ is strictly increasing with $R$, the rate at equilibrium
as a function of $U^{\text{eq}}$ is
\begin{equation}
R_{i}^{\text{e,eq}}=f_{R}^{-1}\left(\boldsymbol{a}_{i}^{\text{eq}},U^{\text{eq}}\right),\mbox{ }i=1,\dots,N,\label{eq:EquilibriumRate-1}
\end{equation}
with $f_{R}^{-1}$ is the inverse of $f$ seen as a function of $R$
only. The value of $U^{\text{eq}}$ is determined from the channel
rate constraint
\begin{equation}
\sum_{i=1}^{N}R_{i}^{\text{e,eq}}=\sum_{i=1}^{N}f_{R}^{-1}\left(\boldsymbol{a}_{i}^{\text{eq}},U^{\text{eq}}\right)=R^{\text{c}}.\label{eq:EqRateConstraint-1}
\end{equation}
 Since $f\left(\boldsymbol{a},R\right)$ is a continuous and strictly
increasing function of $R$, $f_{R}^{-1}\left(\boldsymbol{a},U\right)$
and $\sum_{i=1}^{N}f_{R}^{-1}\left(\boldsymbol{a}_{i},U\right)$ are
also continuous and strictly increasing functions of $U$, with $\sum_{i=1}^{N}f_{R}^{-1}\left(\boldsymbol{a}_{i},0\right)=0$.
Provided that
\begin{equation}
\lim_{U\rightarrow\infty}\sum_{i=1}^{N}f_{R}^{-1}\left(\boldsymbol{a}_{i},U\right)>R^{c},\label{eq:EquilCond-1}
\end{equation}
\eqref{eq:EqRateConstraint-1} admits a unique solution. $\Pi_{i}^{\text{ eq}}$
and $\phi_{i}^{\text{eq}}$,$i=1,\dots,N$, are deduced from~\eqref{eq:Equilibre2}
and~\eqref{eq:EquilibriumRate-1}, provided that $K_{\text{I}}^{\text{e}}\neq0$
and $K_{\text{I}}^{\text{t}}\neq0$.

The equilibrium is thus unique and satisfies the control targets considered
in Section~\ref{sec:problem}. Similar conclusions can be obtained when the buffer
level is controlled.

\subsection{Linearized model}

We study the local stability of the system around
an equilibrium point evaluated in Section~\ref{Ssec:EquilBufferLevel}.
Linearizing \eqref{eq:stateSpace1} around the equilibrium characterized in~\eqref{eq:Equilibre2}
one gets for $i=1,\dots,N$
\begin{equation}
\scalebox{0.8}{$
\left\{ \begin{array}{ll}
\Delta\boldsymbol{a}_{i}\left(j+1\right)& \hspace{-0.3cm}= \Delta\boldsymbol{a}\left(j\right)+\delta\boldsymbol{a}_{i}\left(j\right)\\
\Delta\boldsymbol{a}_{i}^{\text{d}}(j+1) &\hspace{-0.3cm}=\Delta\boldsymbol{a}_{i}(j)\\
\Delta\phi_{i}\left(j+1\right) &\hspace{-0.3cm}=\Delta\phi_{i}\left(j\right)+\frac{1}{N}\sum_{k=1}^{N}\Delta U_{k}^{\text{dd}}\left(j\right)-\Delta U_{i}^{\text{dd}}\left(j\right)\\
\Delta\Pi_{i}^{\tau}\left(j+1\right)&\hspace{-0.3cm}= \Delta\Pi_{i}^{\tau}\left(j\right)-\frac{1}{R_{i}^{\text{e,eq}}}\left(\tau_{0}\Delta\tilde{R_{i}}^{e}(j)-\Delta B_{i}(j)\right)\\
\Delta\tilde{R_{i}}^{\text{e}}(j+1)&\hspace{-0.3cm}=(1-\alpha)\Delta\tilde{R_{i}}^{\text{e}}\left(j\right)+\alpha\Delta R_{i}^{\text{edd}}\left(j\right)\\
\Delta R_{i}^{\text{ed}}(j+1)& \hspace{-0.3cm}= \frac{K_{\text{P}}^{\text{e}}+K_{\text{I}}^{\text{e}}}{T}\frac{1}{R_{i}^{\text{e,eq}}}\left(\tau_{0}\Delta\tilde{R_{i}}^{\text{e}}(j)-\Delta B_{i}(j)\right)-\frac{K_{\text{I}}^{\text{e}}}{T}\Delta\Pi_{i}^{\tau}\left(j\right)\\
\Delta R_{i}^{\text{edd}}(j+1)&\hspace{-0.3cm}= \Delta R_{i}{}^{\text{ed}}(j)\\
\Delta U_{i}^{\text{dd}}\left(j+1\right)&\hspace{-0.3cm}= \frac{\partial f}{\partial\boldsymbol{a}}\left(\boldsymbol{a}_{i}^{\text{d,eq}},R_{i}^{\text{ed,eq}}\right)\Delta\boldsymbol{a}_{i}^{\text{d}}\left(j\right)+\frac{\partial f}{\partial R}\left(\boldsymbol{a}_{i}^{\text{d,eq}},R_{i}^{\text{ed,eq}}\right)\Delta R_{i}^{\text{ed}}\left(j\right)\\
\Delta B_{i}\left(j+1\right)&\hspace{-0.3cm}= \Delta B_{i}\left(j\right)+\Delta R_{i}^{\text{edd}}(j)T\\
&\hspace{-2cm}-  \left(\left(K_{\text{P}}^{\text{t}}+K_{\text{I}}^{\text{t}}\right)\left(\frac{1}{N}\sum_{k=1}^{N}\Delta U_{k}^{\text{dd}}\left(j\right)-\Delta U_{i}^{\text{dd}}\left(j\right)\right)+K_{\text{I}}^{\text{t}}\Delta\phi_{i}\left(j\right)\right)T.
\end{array}\right.\label{eq:EvolLinear}
$}
\end{equation}

Consider the $N\times\left(N\times N^{a}\right)$ block diagonal matrix
\begin{equation}
\Xi=\text{diag}\left(\frac{\partial f}{\partial\boldsymbol{a}^{T}}\left(\boldsymbol{a}_{1}^{\text{d,eq}},R_{1}^{\text{e,eq}}\right),\dots,\frac{\partial f}{\partial\boldsymbol{a}^{T}}\left(\boldsymbol{a}_{N}^{\text{d,eq}},R_{N}^{\text{e,eq}}\right)\right),\label{eq:KsiMatrix-1}
\end{equation}
gathering the sensitivities with respect to $\boldsymbol{a}$ of
the rate-utility characteristics of each stream and the $N\times N$
diagonal matrix
\begin{equation}
\boldsymbol{\Gamma}=\text{diag}\left(\frac{\partial f}{\partial R}\left(\boldsymbol{a}_{1}^{\text{d,eq}},R_{1}^{\text{e,eq}}\right),\dots,\frac{\partial f}{\partial R}\left(\boldsymbol{a}_{N}^{\text{d,eq}},R_{N}^{\text{e,eq}}\right)\right)\label{eq:GammaMatrix-1}
\end{equation}
gathering the sensitivity to $R$ of the rate-utility characteristics
of each stream. Putting all coupled linearized state-space representations
\eqref{eq:EvolLinear} together, one gets a linear discrete-time state-space
representation
\begin{equation}
\begin{array}{l}
\textbf{x}^{\tau}(j+1)=\textbf{A}^{\tau}\textbf{x}^{\tau}(j)+\textbf{w}(j)\end{array}\label{eq:statespace3-2}
\end{equation}
with state vector
\begin{equation}
\scalebox{0.8}{$
\begin{array}{l}
\mathbf{x}^{\tau}=\left(\Delta\boldsymbol{a}, \Delta\boldsymbol{a}^{\text{d}}, \Delta\boldsymbol{\phi}, \Delta\boldsymbol{\Pi}^{\tau}, \Delta\tilde{\boldsymbol{R}}^{\text{e}},\Delta\boldsymbol{R}^{\text{ed}}, \Delta\boldsymbol{R}^{\text{edd}}, \Delta\boldsymbol{U}^{\text{dd}}, \Delta\boldsymbol{B}\right)^{T}\end{array}\label{eq:StateVector}
$}
\end{equation}
and noise input
\begin{equation}
\begin{array}{l}
\mathbf{w}=(\begin{array}{cccc}
\delta\boldsymbol{a} & \mathbf{0} & \dots & \mathbf{0}\end{array})^{T},\end{array}\label{eq:statespace3-1-1}
\end{equation}
representing the fluctuations of the value of the parameter vector
for the rate-utility model. In \eqref{eq:StateVector} and \eqref{eq:statespace3-1-1},
boldface letters represent vectors and time indexes have been omitted. For example $\Delta\boldsymbol{a}\left(j\right)$
is a vector $N\times N^{a}$ components and $\Delta\boldsymbol{B}\left(j\right)=\left(\Delta B_{1}\left(j\right),\dots,\Delta B_{N}\left(j\right)\right)^{T}$
is a vector of $N$ components. From~\eqref{eq:EvolLinear} and~\eqref{eq:statespace3-2},
one deduces
\begin{equation}
\scalebox{0.78}{$
\mathbf{A}^{\tau}=\left(\begin{array}{ccccccccc}
\mathbf{I} & \mathbf{0} & \mathbf{0} & \mathbf{0} & \mathbf{0} & \mathbf{0} & \mathbf{0} & \mathbf{0} & \mathbf{0}\\
\mathbf{I} & \mathbf{0} & \mathbf{0} & \mathbf{0} & \mathbf{0} & \mathbf{0} & \mathbf{0} & \mathbf{0} & \mathbf{0}\\
\mathbf{0} & \mathbf{0} & \mathbf{I} & \mathbf{0} & \mathbf{0} & \mathbf{0} & \mathbf{0} & -\mathbf{L} & \mathbf{0}\\
\mathbf{0} & \mathbf{0} & \mathbf{0} & \mathbf{I} & -\tau_{0}\mathbf{V} & \mathbf{0} & \mathbf{0} & \mathbf{0} & \mathbf{V}\\
\mathbf{0} & \mathbf{0} & \mathbf{0} & \mathbf{0} & (1-\alpha)\mathbf{I} & \mathbf{\mathbf{\mathbf{0}}} & \mathbf{\alpha I} & \mathbf{0} & \mathbf{0}\\
\mathbf{0} & \mathbf{0} & \mathbf{0} & -\frac{K_{\text{I}}^{\tau\text{e}}}{T}\mathbf{I} & \frac{K^{\tau\text{e}}}{T}\tau_{0}\mathbf{\mathbf{V}} & \mathbf{0} & \mathbf{0} & \mathbf{0} & -\frac{K^{\tau\text{e}}}{T}\mathbf{V}\\
\mathbf{0} & \mathbf{0} & \mathbf{0} & \mathbf{0} & \mathbf{0} & \mathbf{I} & \mathbf{0} & \mathbf{0} & \mathbf{0}\\
\mathbf{0} & \boldsymbol{\ensuremath{\Xi}} & \mathbf{0} & \mathbf{0} & \mathbf{0} & \mathbf{\boldsymbol{\Gamma}} & \mathbf{0} & \mathbf{0} & \mathbf{0}\\
\mathbf{0} & \mathbf{0} & -K_{\text{I}}^{\text{t}}T\mathbf{I} & \mathbf{0} & \mathbf{0} & \mathbf{0} & T\mathbf{\mathbf{I}} & K^{\text{t}}T\mathbf{L} & \mathbf{I}
\end{array}\right)
$}
\end{equation}
with $\mathbf{V}=\text{diag}\left(1/R_{1}^{\text{e,eq}},\dots,1/R_{N}^{\text{e,eq}}\right)$
a diagonal matrix containing the inverse of the encoding rates at
equilibrium and $K^{\text{\ensuremath{\tau}e}}=K_{\text{P}}^{\text{\ensuremath{\tau}e}}+K_{\text{I}}^{\text{\ensuremath{\tau}e}}$,
$K^{\text{t}}=K_{\text{P}}^{\text{t}}+K_{\text{I}}^{\text{t}}$. $\mathbf{I}$
and $\mathbf{0}$ are identity and null matrices of appropriate size.

When studying the roots of $\mbox{det}(z\mathbf{I}-\mathbf{A})=0$, $N_{a}\times N$ roots at $z=1$ are obtained. They correspond to
the variations of the rate-utility parameter vector~\eqref{eq:NoiseA}.
The matrix $\boldsymbol{\Xi}$, representing the sensitivity
with respect to $\boldsymbol{a}$ of the rate-utility characteristics
$f$, does not appear in the expressions of $\mathbf{A}^{\tau}$.
Only the sensitivity of $f$ with respect to $R$, represented by
$\boldsymbol{\Gamma}$, impacts the stability around equilibrium.
The system stability is also influenced by the encoding rates at equilibrium
via $\mathbf{V}$ and determined by the PI controller gains $K_{\text{P}}^{\text{t}}$,
$K_{\text{I}}^{\text{t}}$, $K_{\text{P}}^{\text{e}}$, and $K_{\text{I}}^{\text{e}}$.

In Section~\ref{sec:exp}, values of $K_{\text{P}}^{\text{t}}$,
$K_{\text{I}}^{\text{t}}$, $K_{\text{P}}^{\text{e}}$, and $K_{\text{I}}^{\text{e}}$
are chosen so that the system is robust to various realizations
of the rate-utility parameters. The same values of the PI gains are chosen for all
programs. A similar analysis can be done when buffer levels are controlled.

\section{Experimental tests}
\label{sec:exp}

\subsection{Example of application context}
\label{sub:application}

A typical application scenario for the proposed rate control system
is Mobile TV using the evolved MBMS standard
\cite{ETSI_MBMS_03}. The MBMS architecture is composed of three main
entities: BM-SC, MBMS-GW and MCE. The Multicast/Broadcast Service
Center (BM-SC) is a node that serves as entry point for the content
providers delivering the video sources, used for service announcements,
session management. The MANE, considered in the paper in charge of
choosing the encoding and the transmission rates, may be located at
the Broadcast/Multicast source at the entrance of the BM-SC node.
The MBMS-Gateway (GW) is an entity responsible for distributing the
traffic across the different eNBs belonging to the same broadcast
area. It ensures that the same content is sent from all the eNBs by
using IP Multicast. The Multi-cell/multicast Coordination Entity (MCE)
is a logical entity, responsible for allocation of time and frequency
resources for multi-cell MBMS transmission. As in~\cite{Vukadinovic2008},
we assume that the MBMS-GW periodically notifies the MCE about
the resource requirements of video streams so that the resources at
eNBs can be re-allocated accordingly. Therefore, the BM-SC should
ensure that the encoding rate of the multiplex does not violate the
already allocated resources. This is obtained thanks to the proposed
rate control scheme.

\subsection{Simulation environment}

To illustrate the properties of the proposed controllers, this section
describes a simulation of mobile TV delivery in the previously described context.
We consider $N=6$ video streams, each of $100$~s long, extracted
from real TV programs. Interview\footnote{http://www.youtube.com/watch?v=l2Y5nIbvHLs} (Prog 1), Sport\footnote{http://www.youtube.com/watch?v=G63TOHluqno} (Prog~2), Big Buck Bunny\footnote{http://www.youtube.com/watch?v=YE7VzlLtp-4}
(Prog~3), Nature Documentary\footnote{http://www.youtube.com/watch?v=NNGDj9IeAuI} (Prog~4), Video Clip\footnote{http://www.youtube.com/watch?v=rYEDA3JcQqw} (Prog~5), and
an extract of \emph{Spiderman}\footnote{http://www.youtube.com/watch?v=SYFFVxcRDbQ} (Prog~6) in 4CIF ($704\times576$)
format are encoded with x.264~\cite{HomePage2011} at a frame rate
$F=30$~fps. GoPs of $10$~frames are considered, thus the GoP
duration is $T=0.33$~s. The videos, already encoded using MPEG-4,
have been converted to YUV format using \emph{ffmpeg}~\cite{ffmpeg}.
The average rate and PSNR of the streams encoded by x.264 with a constant
quantization parameter $QP=3$ are provided in Table~\ref{tab:videos}.
\begin{table}[htpb]
 \centering %
\begin{tabular}{|c|c|c|c|}
\hline
Video  & Rate (kbit/s)  & PSNR (dB)  & Activity \tabularnewline
\hline
Prog 1  & 1669.9  & 46.06  & low \tabularnewline
\hline
Prog 2  & 4929.1  & 44.23  & high \tabularnewline
\hline
Prog 3  & 3654.6  & 44.56  & high \tabularnewline
\hline
Prog 4  & 2215.1  & 44.61  & low \tabularnewline
\hline
Prog 5  & 2811.4  & 46.37  & medium \tabularnewline
\hline
Prog 6  & 3315.9  & 46.53  & high \tabularnewline
\hline
\end{tabular}\caption{Average rate and PSNR of the six considered video streams (encoding
with x.264 and constant $QP=3$.}
\label{tab:videos}
\end{table}

The videos are then processed with the proposed control system operating
at the GoP level. Initially, all buffers contain
three encoded GoPs corresponding to a buffering delay of $1$~s.
The size $B_{\max}$ of the buffers is taken large enough to support
the variations of its level, occurring, \emph{e.g.}, during scene
changes. Here, their size in bits is $B_{\max}=4$~Mbits. The reference
buffer level in bits is $B_{0}=400$~kbits and the reference delay
is $\tau_{0}=1.5$~s. This reference delay is consistent with a typical
switching time of less than $2$~s, as expected in MBMS Television
services \cite{ARIB2008}. The channel rate is $R^{\text{c}}=4$~Mbps.
The encoding rates are initially considered equal to $R_{0}=R^{\text{c}}/N$.
These rates correspond to the output value of the rate control process
provided to each video server. The encoder is then in charge of adjusting
its encoding parameters to achieve the target bit rate.

The encoding/transcoding rates are sent to the
video encoders which have to choose the encoding parameters for the
next VU. In the considered simulation, the video quality is in terms of
PSNR or of SSIM of the encoded VUs. This quality metric
is transmitted to the MANE in the packet headers. Note that the utility
model~\eqref{Eq:psnr2} is only required to characterize the stability
of the system and to tune the control parameters. Once the parameters
have been chosen off-line, there is no need to know precisely the
model~\eqref{Eq:psnr2} within the MANE.

The proposed \emph{quality fair} (QF) video delivery system is compared
to a \emph{transmission rate fair} (TRF) controller which provides
equal transmission rate to the $N$ video streams. In the TRF
scheme, the encoding rate is controlled to limit the buffer level/delay discrepancy.

A comparison is also performed with a \emph{utility max-min fair} (UMMF) approach \cite{Cho2005},
with a proportional transmission rate control limiting the buffer
level discrepancy. In the UMMF approach, the MANE tries to find the set
of encoding rates for the next VU that maximizes the minimum utility.
The following constrained optimization problem is then considered
\begin{equation}
\begin{array}{l}
\mathbf{R^e}\left(j+1\right) = \\
\arg \smash{\displaystyle\max_{R^e_{1},\dots,R_{N}}}\min\left\{ f\left(\mathbf{a}_{1}(j),R^e_{1}\right),\dots,f\left(\mathbf{a}_{N}(j),R^e_{N}\right)\right\}
\label{eq:Utilitymaxmin}
\end{array}
\end{equation}
\[
\text{such that} \sum_{i=1}^{N}R^e_{i}=R^{c}.
\]

Solving~\eqref{eq:Utilitymaxmin} requires the availability at the
MANE of all rate-utility characteristics (or at least all vectors of
parameters $\mathbf{a}_{i}(j)$) of the previously encoded VUs, contrary
to the QF approach, where only the actual utility of the VUs is needed.
The fact that $\mathbf{a}_{i}(j)$ is used in (\ref{eq:Utilitymaxmin})
for the evaluation of the encoding rate at time $j+1$ accounts for the
possibility for the MANE to get only rate-utility characteristics of previously encoded and already received VUs.
Once the value of the encoding rates $R^e_{1},\dots,R_{N}$ are derived, a proportional (P) controller for the transmission rate is applied to evaluate the transmission rate allocated to each video stream
\begin{equation}
R_{i}^{\text{t}}\left(j\right)=R_{0}+K_{P}^{\text{t}}(B_i(j)-B_0),\label{eq:transmission2}
\end{equation}
where $K_{\text{P}}^{\text{t}}$ is the proportional correction gain.

%An other possible comparison case can be considered corresponding to a fully centralized solution that formulated into a constrained optimization system that maximize the quality under fairness, bandwidth and buffer level constraints. However, this solution is much more complex than the proposed PI-based one since it requires the R-D models of each VU in each multiplexed stream, which increases the computation complexity and the signaling between the $N$ servers and the MANE. Also, the proposed solution allows for a more scalable system than Centralized Optimized System. Using PI controllers allows for multiplexing more flows without highly increasing the complexity. The only centralized operation is the calculation of the average video quality.
%In the simulation part, for fair comparison, we provide results of the QF and TRF solutions.
In this section different cases are considered: Both buffer level and buffering delay are addressed separately including stability analysis
and results for different utility metrics. Then, the robustness of the proposed control system is analyzed by considering variations of the channel rate as well as of the number of video programs.

\subsection{Control of the buffer level}

We first focus on the system performance when the buffer level (in bits) is used
to update the encoding rate.

\subsubsection{PSNR-rate model}

To tune the PI controllers of the QF system, the first utility function considered
is the PSNR of each GoP. As in~\cite{Zhu2007}, a logarithmic PSNR-rate
model is used
\begin{eqnarray}
U_{i}(j)&=P_{i}(j)=f\left(\boldsymbol{a}_{i}\left(j\right),R_{i}^{\text{e}}(j)\right)\\ \nonumber
&=a_{i}^{(1)}(j)\log(a_{i}^{(2)}(j)R_{i}^{\text{e}}(j)),\label{eq:model}
\end{eqnarray}
with $P_{i}(j)$ the PSNR of the GoP at time $j$ for the
$i$-th stream. For the $N=6$ considered programs, the entries of
$\boldsymbol{a}_{i}\left(j\right)$ are estimated for each GoP using
four encoding trials.
\begin{figure}[h!]
\centering \includegraphics[width=\columnwidth]{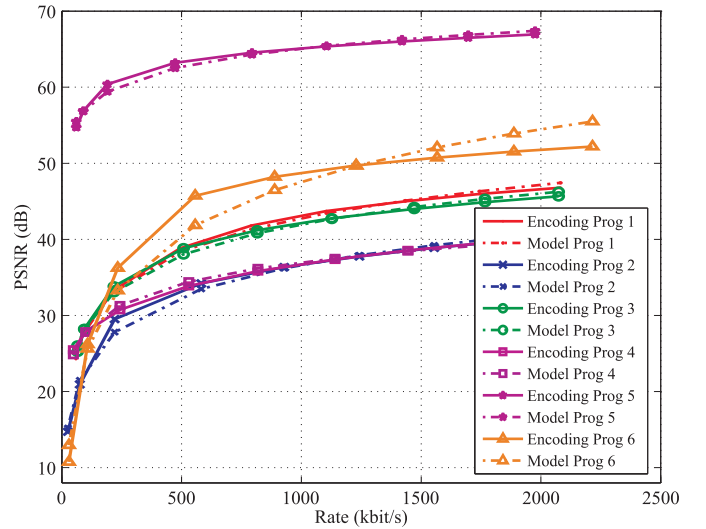}
\caption{PSNR-rate characteristics and models for the first GoP of the $N=6$
considered programs}
\label{fig:model}
\end{figure}
An example of the accuracy of the PSNR-rate
model~\eqref{eq:model} is shown in Figure~\ref{fig:model}, when
applied on the first GoP of the six considered programs, with parameters
estimated from encoding trials performed at 80~kb/s,
200~kb/s, 800~kb/s, and 2~Mb/s.
Figure~\ref{fig:model} illustrates
the ability of~\eqref{eq:model} to predict the PSNR over a wide range
of rates. In addition, the correlation coefficient $r^{2}$ between
experimental and predicted PSNR-rate points is evaluated as
\begin{equation}
r^{2}=\frac{\sigma_{xy}^{2}}{\sigma_{x}^{2}\sigma_{y}^{2}}\label{eq:model-1}
\end{equation}
with
%\begin{eqnarray}
%\sigma_{x}^{2}&=&\sum_{k=1}^{n}(x_{k}-\bar{x}){}^{2},\\ \sigma_{y}^{2}&=&\sum_{k=1}^{n}(y_{k}-\bar{y}){}^{2},\\
%\sigma_{xy}^{2}&=&\sum_{k=1}^{n}(y_{k}-\bar{y})(x_{k}-\bar{x}),
%\end{eqnarray}
$\sigma_{x}^{2}=\sum_{k=1}^{n}(x_{k}-\bar{x}){}^{2}$, $\sigma_{y}^{2}=\sum_{k=1}^{n}(y_{k}-\bar{y}){}^{2}$, and
$\sigma_{xy}^{2}=\sum_{k=1}^{n}(y_{k}-\bar{y})(x_{k}-\bar{x})$,
where $n$ is the number of rates for each program at which the PSNR
has been evaluated ($x_{k}$) and predicted ($y_{k}$) using~\eqref{eq:model},
and where $\bar{x}$ and $\bar{y}$ are the average values of the
$x_{k}$'s and of the $y_{k}$'s. For the six programs, for $n=7$, the
rate values are 80~kb/s, 130~kb/s, 200~kb/s, 500~kb/s, 800~kb/s,
1.4~Mb/s, and 2~Mb/s, the correlation coefficients are $r^{2}=[0.998,0.996,0.997,0.996,0.992,0.985]$
illustrating to good fit by~\eqref{eq:model} of the PSNR-rate characteristics.

\subsubsection{Controller design and stability analysis}
\label{Ssec:StabPSNRBufferLevel}

The values of the parameter vector $\boldsymbol{a}_{i}\left(1\right)$,
$i=1,\dots,6$, obtained for the first GoP of the $N=6$ programs are
\begin{equation}
\scalebox{0.9}{$
\begin{array}{ll}
\boldsymbol{a}_{1}\left(1\right) & =\left(\begin{array}{c}
1.11\\
0.15
\end{array}\right),\,\boldsymbol{a}_{2}\left(1\right)=\left(\begin{array}{c}
1.90\\
0.17
\end{array}\right),\,\boldsymbol{a}_{3}\left(1\right)=\left(\begin{array}{c}
0.76\\
0.17
\end{array}\right),\nonumber \\
\boldsymbol{a}_{4}\left(1\right) & =\left(\begin{array}{c}
0.09\\
0.24
\end{array}\right),\,\boldsymbol{a}_{5}\left(1\right)=\left(\begin{array}{c}
2.50\\
0.17
\end{array}\right),\,\boldsymbol{a}_{6}\left(1\right)=\left(\begin{array}{c}
0.07\\
0.20
\end{array}\right).\label{eq:a_parameters}
\end{array}
$}
\end{equation}
Once $f$ is specified, one may characterize the system equilibrium.
The vector of rates at equilibrium $\left(R_{1}^{\text{e,eq}},\dots,R_{N}^{\text{e,eq}}\right)^{T}$
is obtained by solving the system of equations in the state-space representation at equilibrium. $\boldsymbol{\Xi}$ and $\boldsymbol{\Gamma}$ are derived from~\eqref{eq:KsiMatrix-1}, \eqref{eq:GammaMatrix-1} and~\eqref{eq:model} as follows
\begin{equation}
\scalebox{0.6}{$
\boldsymbol{\Xi}=\left(\begin{array}{ccccccc}
\log(a_{1,2}\left(1\right)R_{1}^{\text{e,eq}}) & \frac{a_{1,1}\left(1\right)}{a_{1,2}\left(1\right)} & 0 & 0 & \dots & 0 & 0\\
0 & 0 & \log(a_{2,2}\left(1\right)R_{2}^{\text{e,eq}}) & \frac{a_{2,1}\left(1\right)}{a_{2,2}\left(1\right)} & 0 & \dots & 0\\
 &  & 0 & 0 & \ddots &  & 0\\
\vdots & \vdots & \vdots & \ddots &  & \ddots & 0\\
0 & 0 & 0 & 0 & \dots & \log(a_{N,2}\left(1\right)R_{N}^{\text{e,eq}}) & \frac{a_{N,1}\left(1\right)}{a_{N,2}\left(1\right)}
\end{array}\right)\label{eq:xiPSNR}
$}
\end{equation}
and
\begin{equation}
\boldsymbol{\Gamma}=\text{diag}\left(\frac{a_{1,1}\left(1\right)}{R_{1}^{\text{e,eq}}\left(1\right)},\dots,\frac{a_{N,1}\left(1\right)}{R_{N}^{\text{e,eq}}\left(1\right)}\right).\label{eq:gammaPSNR}
\end{equation}

The gains of the PI controllers have to be chosen so that the roots of
\begin{equation}
d(z)=\det\left(z\mathbf{I}-\mathbf{A}\right)
\label{eq:CharPol}
\end{equation}
remains within the unit circle,
for various rate-utility characteristics of the VUs.
In \eqref{eq:CharPol}, $\mathbf{A}$ may correspond to $\mathbf{A}^{b}$, the linearized state matrix when considering buffer level control, or to $\mathbf{A}^{\tau}$ with buffering delay control.

\begin{figure}
\centering \includegraphics[width=\columnwidth]{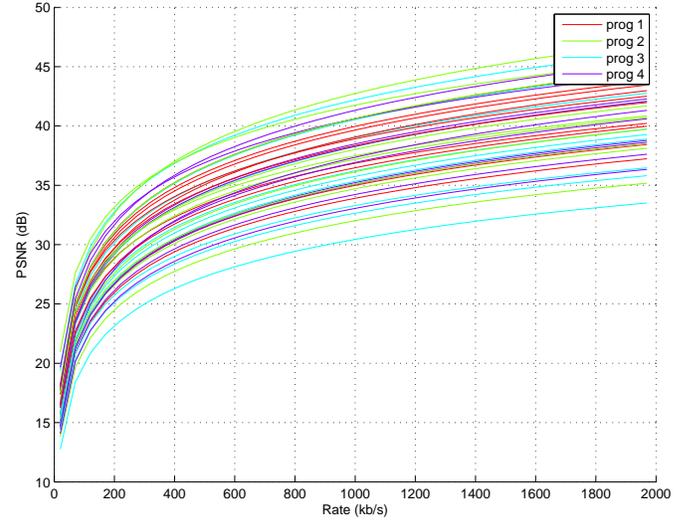}
\caption{Superposition of the $K=10$ realizations of the $N=4$ random PSNR-rate
characteristics}
\label{fig:video}
\end{figure}

To increase the robustness of the proposed approach to variations of the rate-utility characteristics,
$K=10$ realizations of $N=4$ random parameter vectors of the PSNR-rate model obtained as follows
\begin{eqnarray}
a_{i,1}^{(k)}&=\frac{1}{N}\sum_{i=1}^{N}a_{i,1}\left(1\right)+\eta_{i,1}^{(k)},\\\nonumber
a_{i,2}^{(k)}&=\frac{1}{N}\sum_{i=1}^{N}a_{i,2}\left(1\right)+\eta_{i,2}^{(k)},\label{eq:random}
\end{eqnarray}
for $k=1,\dots,K$. In~\eqref{eq:random}, $\eta_{i,1}^{(k)}$ and $\eta_{i,2}^{(k)}$
are realizations of zero-mean Gaussian variables with variance $\sigma_{1}^{2}=6.25\times10^{-2}$
and $\sigma_{2}^{2}=2.25\times10^{-4}$. The resulting PSNR-rate characteristics
obtained using~\eqref{eq:random} are represented in Figure~\ref{fig:video}.
These random realizations describe quite well the variability of actual
PSNR-rate characteristics represented in Figure~\ref{fig:model}.

A random search of the control parameters is then performed. Among the values providing stability
for the $K$ random PSNR-rate characteristics, the one with the roots farthest away from the unit circle is selected to provide good transients.

\begin{figure}
\centering \includegraphics[width=\columnwidth]{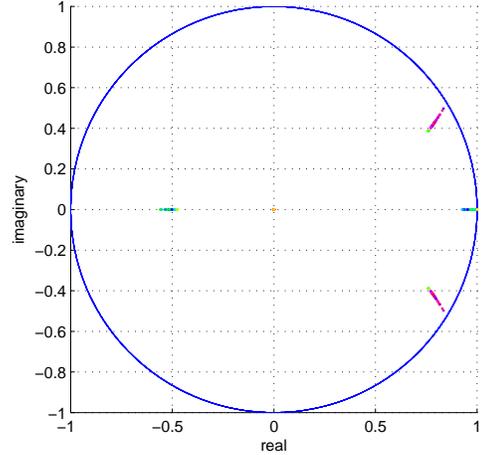}
\caption{Location of the roots of $\det\left(z\mathbf{I}-\mathbf{A}^{\text{b}}\right)$
for the $K=10$ realizations of the $N=4$ random PSNR-rate characteristics}
\label{fig:roots1}
\end{figure}

The tuning is performed for $N=4$. Good
transient behaviors have been obtained with $K_{\text{P}}^{\text{e,b}}=666$,
$K_{\text{I}}^{\text{e,b}}=33$, $K_{\text{P}}^{\text{t}}=66\times10^{3}$,
and $K_{\text{I}}^{\text{t}}=1300$.
The position of the roots corresponding to the video characteristic
represented in Figure~\ref{fig:video} are in Figure~\ref{fig:roots1}.

Figure~\ref{fig:roots1} shows that all roots remain within
the unit circle. This result does not prove the robustness of the proposed
choice of the control parameters, but shows that this choice leads to a system
reasonably robust to changes of the characteristics
of the transmitted programs. Nevertheless, some of the roots are located quite near
the stability limit, which will lead to quite long transients.

\subsubsection{Simulation results}

The x.264 video encoder is used to perform on-line compression of the
different programs with the rate targets provided by the encoding rate
controllers. Within the video coder, a two-pass
rate control is performed to better fit the target encoding rate.
The target and obtained encoding rates may however be slightly different.
The system performance is first measured in terms of average buffer
level discrepancy $\Delta_{B}$ (in bits) with respect to $B_{0}$, variance of the buffer level $\sigma_{B}^{2}$ (in bits$^{2}$),
PSNR discrepancy $\Delta_{P}$ (in dB), and PSNR variance $\sigma_{P}^{2}$
(in dB$^{2}$), with
\begin{equation}
\begin{array}{l}
\Delta_{B}  =  \frac{1}{NM}\sum_{n=1}^{N}\sum_{l=1}^{M}\left(B_{n}(l)-B_{0}\right),\\
\sigma_{B}^{2}  = \frac{1}{N}\sum_{n=1}^{NM}\sum_{l=1}^{M}\left(B_{n}(l)-B_{0}-\Delta_{B}\right)^{2},\\
\Delta_{PSNR}  =  \frac{1}{NM}\sum_{n=1}^{N}\sum_{l=1}^{M}\left(P_{n}(l)-\bar{P}(l)\right),\\
\sigma_{PSNR}^{2}  =  \frac{1}{N}\sum_{n=1}^{NM}\sum_{l=1}^{M}\left(P_{n}(l)-\bar{P}(l)-\Delta_{PSNR}\right)^{2},
\end{array}\label{eq:sys_perf}
\end{equation}
where $\bar{P}(l)=\frac{1}{N}\sum_{n=1}^{N}P_{n}(l)$ and $M$ is
the number of GoPs in the video streams.

The results with $N=6$ and $R^{\text{c}}=4$~Mbit/s are summarized
in Table~\ref{tab:control_bits_video} in the three cases: TRF, UMMF, and
QF where $\Delta_{B}$ are in kbits and $\sigma_{B}^2$ in kbit$^2$. The PI controllers used for the transmission rate control loop
reduce the PSNR discrepancy between the programs at a price of some
increase of the buffer level discrepancy and variance.
\begin{table}[h]
\centering %
\begin{tabular}{|c|c|c|c|c|c|c|}
\hline
 & $K_{\text{P}}^{\text{e,b}},K_{\text{I}}^{\text{e,b}}$  & $K_{\text{P}}^{t},K_{\text{I}}^{t}$  & $\left|\Delta_{B}\right|$  & $\sigma_{B}^{2}$  & $\left|\Delta_{P}\right|$  & $\sigma_{P}^{2}$ \tabularnewline
\hline
TRF  & $666,0$  & $0,0$  & $31.5$  & $2.1$  & $3.1$  & $9.8$ \tabularnewline
\hline
UMMF  &  $0,0$  & $3,0$  & $75$  & $2$  & $2.7$  & $13.9$ \tabularnewline
\hline
QF  & $666,33$  & $(66,1.3)10^3$  & $53.07$  & $6.1$  & $1.5$  & $6.7$ \tabularnewline
\hline
\end{tabular}\caption{Performance when using TRF and QF controllers when controlling the
buffer levels for $N=6$.}
\label{tab:control_bits_video}
\end{table}

\begin{figure}[h!]
\centering \includegraphics[width=\columnwidth]{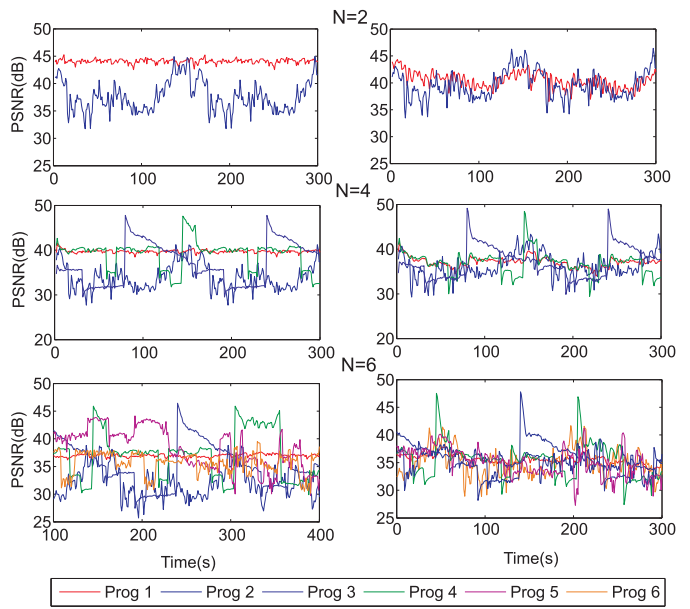}
\caption{Evolution of the PSNR for $N=2$, $N=4$, and $N=6$ using
TRF (left) and QF (right) controllers when controlling the buffer
levels}
\label{fig:psnr_bits}
\end{figure}

Figure~\ref{fig:psnr_bits} represents the evolution of the PSNR
of the $N$ programs over $300$ GoPs using the TRF (left) and the
QF (right) controllers when $N=2$, $N=4$, and $N=6$ with a constant
channel rate $R^{\text{c}}=4$~Mbit/s.
The same controller parameters
$K_{\text{P}}^{\text{e,b}}=666$, $K_{\text{I}}^{\text{e,b}}=33$,
$K_{\text{P}}^{\text{t}}=66\times10^{3}$, and $K_{\text{I}}^{\text{t}}=1300$
are used for $N=2$, $N=4$, and $N=6$.

When $N=2$, the average PSNR of Prog $2$, characterized by high
activity level, is improved from $36$~dB to $41$~dB, leading to
a significant improvement of the video quality. This is at the price
of PSNR degradation of Prog~$1$, characterized by low activity level,
from $45$ dB to $41$ dB. This still corresponds to a very good quality.
PSNR fairness improvements are also obtained when $N=4$ and $N=6$.

\begin{figure}[h]
\centering \includegraphics[width=\columnwidth]{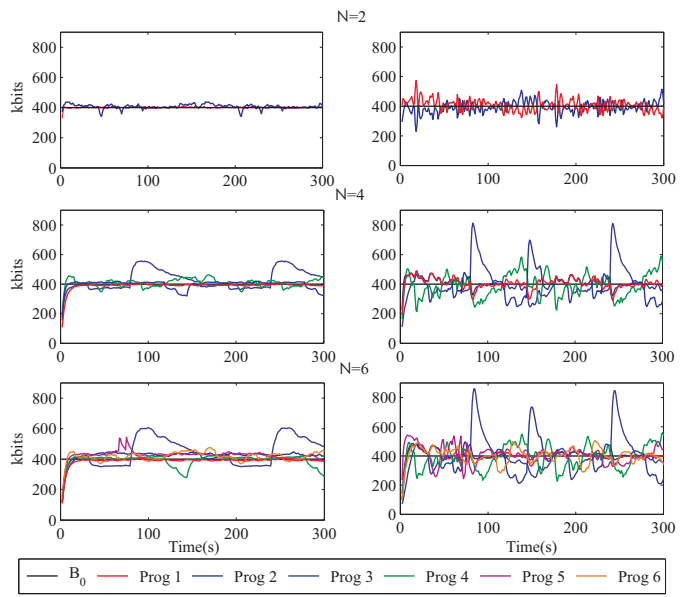}
\caption{Evolution of the buffer level for $N=2$, $N=4$, and $N=6$
using TRF (left) and QF (right) controllers when controlling the buffer
levels}
\label{fig:buffer_bits}
\end{figure}

Figure~\ref{fig:buffer_bits} shows the evolution of the buffer level
of the $N$ programs over $300$ GoPs using the TRF (left) and the
QF (right) controllers for $N=2$, $N=4$, and $N=6$.
The discrepancy between the buffer level and the reference level $B_{0}$ remains
limited for most of the time. When only the encoding rate is controlled,
corresponding to the TRF controller, the buffer level stabilizes around
$B_{0}$.
The buffer level variations increase using the QF controller
due to the interactions of the encoding rate and transmission rate
control loops.

\begin{figure}[h]
\centering
\includegraphics[width=\columnwidth]{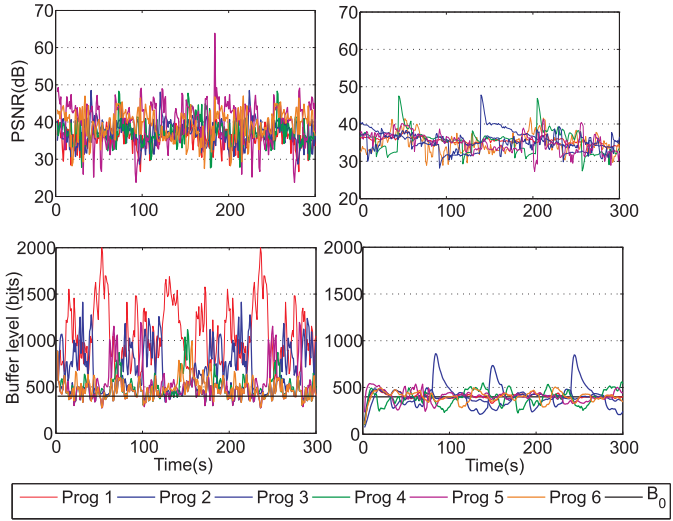}
\caption{Evolution of the buffer level and of the PSNR (right) for $N=6$
using the UMMF technique (left) and using the QF controllers (right) with a transmission rate control using the buffer levels}
\label{fig:UMMF}
\end{figure}

Figure~\ref{fig:UMMF} shows the evolution of the buffer level (left) and of the PSNR (right)
of the $N$ programs over $300$ GoPs when the UMMF technique is used. The choice of the proportional gain for the transmission rate controller has no significant impact on the quality fairness, provided that the buffers remain full. A reference buffer level
of $400$~kbits has been used to allow a satisfying behavior of the transmission rate control loop.
The PSNR remains around $40$~dB, but the average variance is of the same order of magnitude as that of the
TRF solution, see Table~\ref{tab:control_bits_video}. This mainly comes from the target encoding rate evaluation on (one step) outdated PSNR-rate characteristics.

\subsubsection{SSIM-rate model}

The quality fairness is also addressed considering the SSIM metric.
To tune the PI controllers, an arctan SSIM-Rate utility model is considered
\begin{eqnarray}
U_{i}(j)&=S_{i}(j)=f\left(\boldsymbol{a}_{i}\left(j\right),R_{i}^{\text{e}}(j)\right)\\\nonumber
&=a_{i}^{(1)}(j)\text{{atan}}(a_{i}^{(2)}(j)R_{i}^{\text{e}}(j))\label{eq:model2}
\end{eqnarray}
where $S_{i}(j)$ is the SSIM of program $i$ at time $j$. As before,
the two entries of each $\boldsymbol{a}_{i}\left(j\right)$ are derived
from four encoding trials performed on each of the $N$ considered
programs. The resulting values of the parameters are
\begin{equation}
\scalebox{0.9}{$
\begin{array}{l}
\boldsymbol{a}_{1}\left(1\right) =\left(\begin{array}{c}
0.64\\
0.037
\end{array}\right),\,\boldsymbol{a}_{2}\left(1\right)=\left(\begin{array}{c}
0.61\\
0.029
\end{array}\right),\,\boldsymbol{a}_{3}\left(1\right)=\left(\begin{array}{c}
0.64\\
0.034
\end{array}\right),\nonumber \\
\boldsymbol{a}_{4}\left(1\right) =\left(\begin{array}{c}
0.62\\
0.017
\end{array}\right),\,\boldsymbol{a}_{5}\left(1\right)=\left(\begin{array}{c}
0.64\\
0.22
\end{array}\right),\,\boldsymbol{a}_{6}\left(1\right)=\left(\begin{array}{c}
0.64\\
0.044
\end{array}\right).\label{eq:a_parameters-1}
\end{array}
$}
\end{equation}

\begin{figure}[htp]
\centering
\includegraphics[width=\columnwidth]{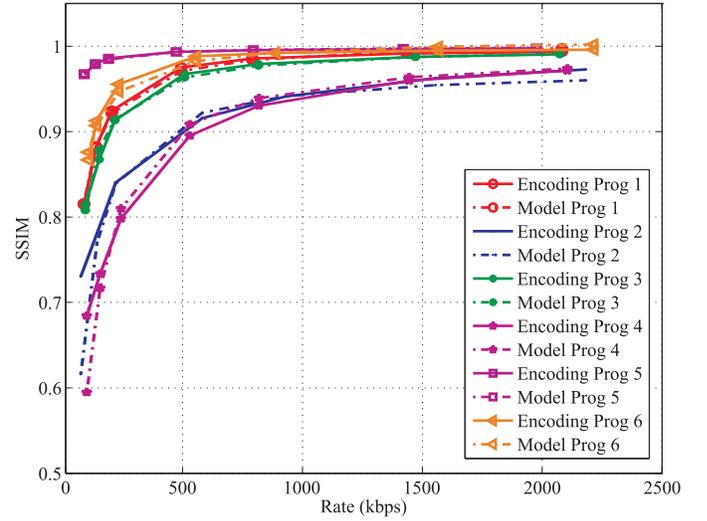}
\caption{SSIM-Rate model for the six considered programs}
\label{fig:model2}
\end{figure}

Figure~\ref{fig:model2} compares the actual SSIM-Rate characteristics
and those obtained using the model~\eqref{eq:model2} for the first
GoP of the six considered programs. The model~\eqref{eq:model2}
is able to predict accurately the SSIM over a large range of rates.
The correlation coefficient for the six programs is $r^{2}=[0.99,0.97,0.99,0.98,0.99,0.99]$
using the same rate values in the PSNR-rate model estimation, which
confirms the accuracy of the SSIM-rate model.

Using the SSIM-rate utility function~\eqref{eq:model2}, one is able to get the vector
$\left(R_{1}^{\text{e,eq}},\dots,R_{N}^{\text{e,eq}}\right)^{T}$
of encoding rates as the solution of \eqref{eq:resolution-1}.
$\boldsymbol{\Xi}$ and $\boldsymbol{\Gamma}$ are derived from~\eqref{eq:KsiMatrix-1}, \eqref{eq:GammaMatrix-1} and~\eqref{eq:model2} as follow
\begin{equation}
\scalebox{0.78}{$
\boldsymbol{\Xi}=\left(\begin{array}{ccccc}
\text{atan}(a_{1}^{(2)}R_{1}^{\text{e,eq}}) & \frac{a_{1}^{(1)}R_{1}^{\text{e,eq}}}{1+(a_{1}^{(2)}R_{1}^{\text{e,eq}})^{2}} & 0 & 0 & 0\\
0 & \ddots & \ddots & \vdots & \vdots\\
\vdots & 0 & \ddots & 0 & 0\\
0 & \dots & 0 & \text{atan}(a_{N}^{(2)}R_{N}^{\text{e,eq}}) & \frac{a_{N}^{(1)}R_{N}^{\text{e,eq}}}{1+(a_{N}^{(2)}R_{N}^{\text{e,eq}})^{2}}
\end{array}\right)\label{eq:xiSSIM}
$}
\end{equation}
and
\begin{equation}
\boldsymbol{\Gamma}=\text{diag}\left(\begin{array}{ccc}
\frac{a_{1}^{(1)}a_{1}^{(2)}}{1+(a_{1}^{(2)}R_{1}^{\text{e,eq}})^{2}} & \dots & \frac{a_{N}^{(1)}a_{N}^{(2)}}{1+(a_{N}^{(2)}R_{N}^{\text{e,eq}})^{2}}\end{array}\right)\label{eq:gammaSSIM}
\end{equation}

The choice of the parameters of the controllers is done as in Section~\ref{Ssec:StabPSNRBufferLevel}.
Good transient behaviors have been obtained with $K_{\text{P}}^{\text{e,b}}=666$,
$K_{\text{I}}^{\text{e,b}}=33$, $K_{\text{P}}^{\text{t}}=66\times10^{4}$,
and $K_{\text{I}}^{\text{t}}=1.3\times10^{4}$.
\begin{figure*}[htpb]
\centering \includegraphics[width=0.9\columnwidth]{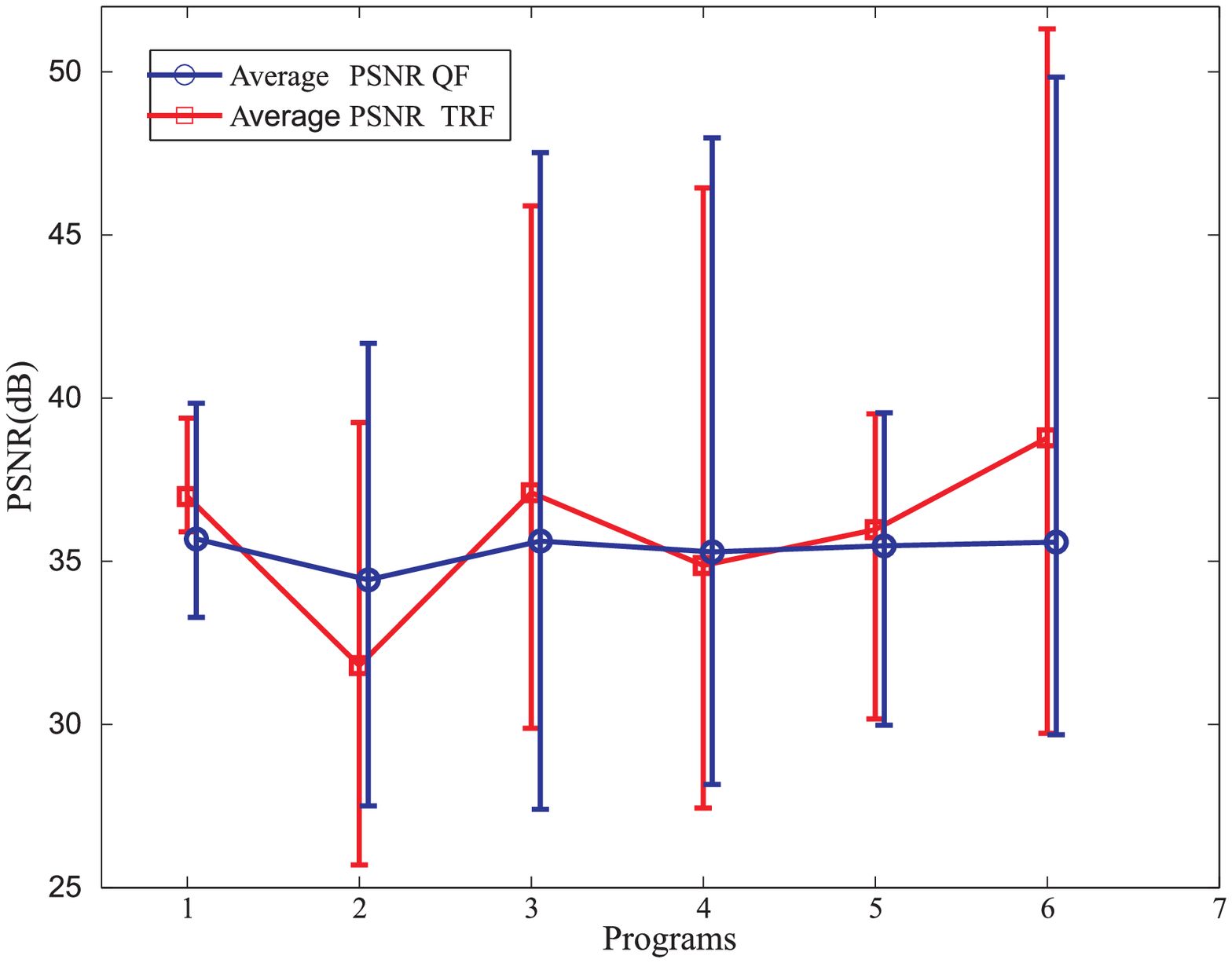}
\includegraphics[width=0.92\columnwidth]{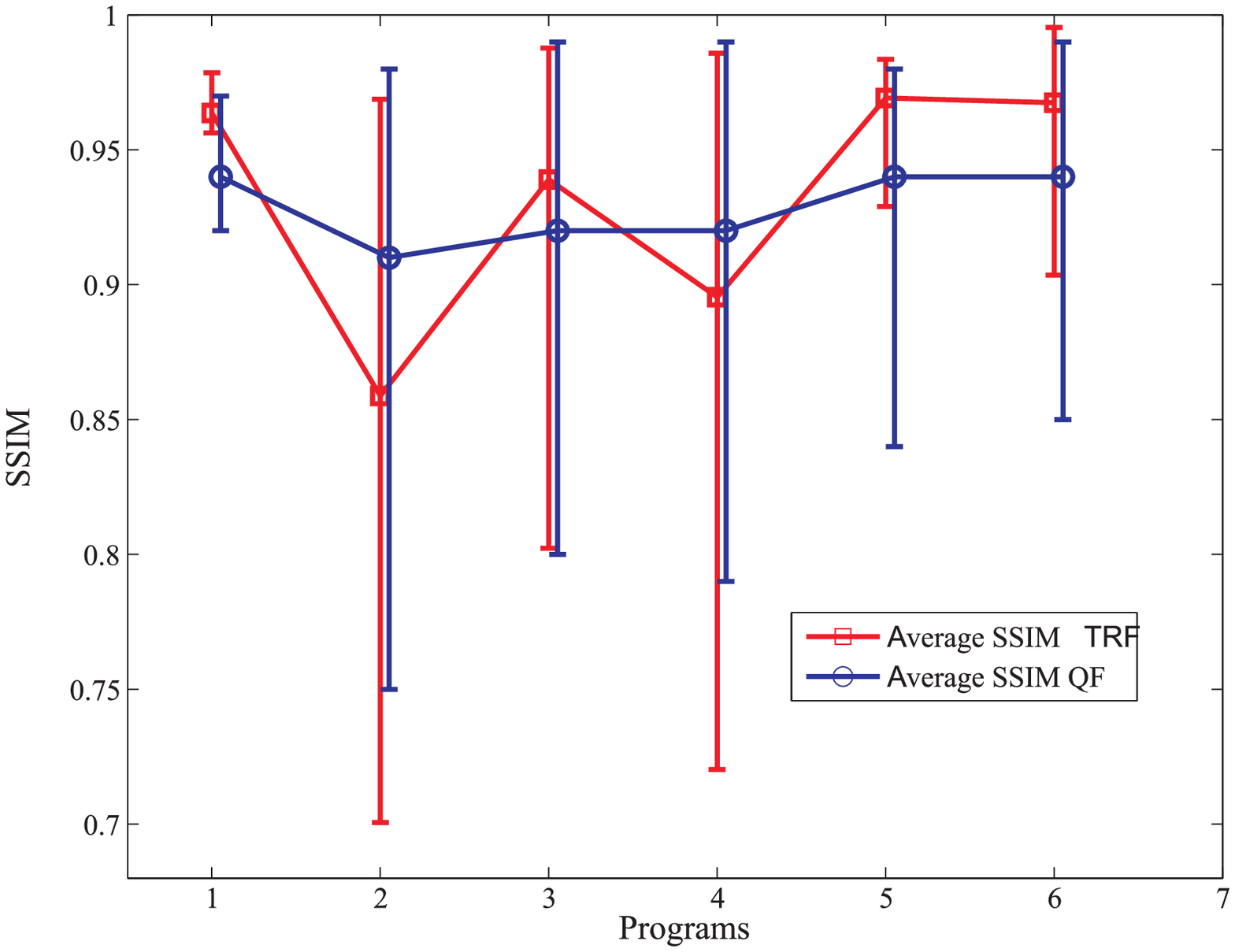}
\caption{Minimum, average, and maximum values of the PSNR (left) and SSIM (right)
over all GoPs for the TRF and the QF controllers using $N=6$ programs
when controlling the buffer levels}
\label{fig:stat}
\end{figure*}
For this choice of the control gains, Figure~\ref{fig:stat} shows the minimum, average,
and maximum values of the PSNR (left) and SSIM (right) over all GoPs
of the $N=6$ programs. The QF controller improves quality fairness
especially for the most demanding videos, such as Prog~$2$. The proposed
QF controller improves also the minimum achieved quality for these
programs. In fact, even if events corresponding to these minimum quality
happens only few times, the user perception is sometime dominated
by the worst experience, rather than the average. The price to be
paid is, as expected, a decrease of the quality of the less demanding
programs.

\subsection{Control of the buffering delays}
\label{Ssec:BufferingDelayControlEx}

This part focuses on the system performance when
the buffering delays are used to evaluate the encoding rates.

\subsubsection{Controller design}

The same PSNR-rate utility model as in~\eqref{eq:model} is considered
in this section. Thus the matrices $\Xi$ and $\boldsymbol{\Gamma}$
are those in~\eqref{eq:xiPSNR} and~\eqref{eq:gammaPSNR}.

The choice of the parameters of the two PI controllers is again done as in Section~\ref{Ssec:StabPSNRBufferLevel}. Now, the
roots of $\det\left(z\mathbf{I}-\mathbf{A}^{\tau}\right)$ have to remain within
the unit circle. Good transient behaviors have been obtained for $N=2$, $N=4$, and $N=6$
with $K_{\text{P}}^{\text{t}}=66\times10^{3}$, $K_{\text{I}}^{\text{t}}=2600$
$K_{\text{P}}^{\text{e}\tau}=66\times10^{3}$, and $K_{\text{I}}^{\text{e}\tau}=1300$.

\begin{figure}[h]
\centering \includegraphics[width=\columnwidth]{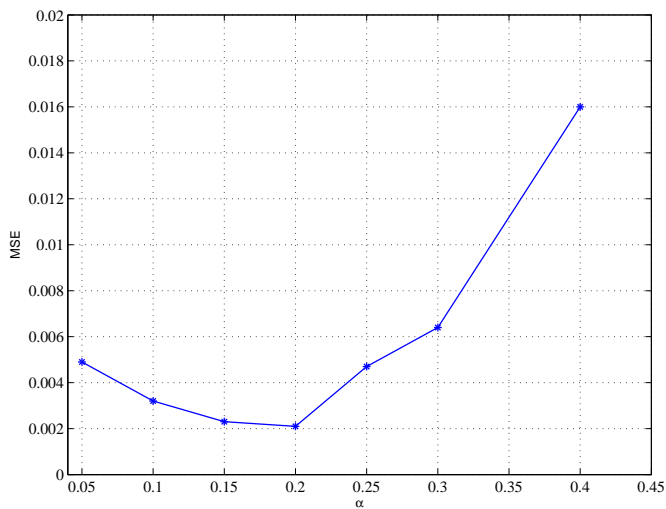}
\caption{MSE of the estimated buffering delay as a function of $\alpha$.}
\label{fig:alpha}
\end{figure}

\begin{figure}[t!h]
\centering \includegraphics[width=\columnwidth]{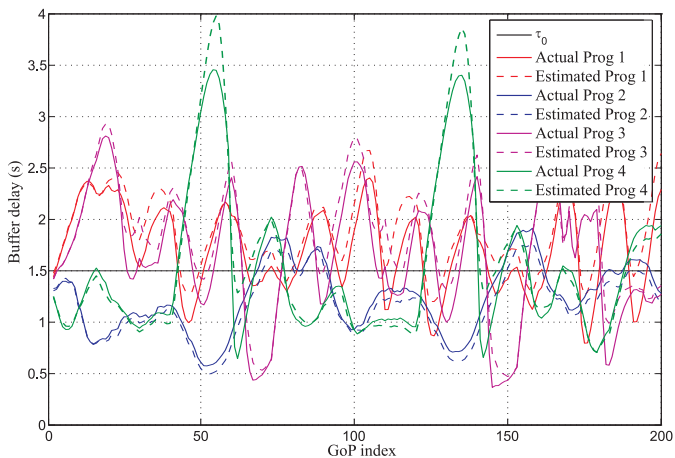}
\caption{Evolution of the actual buffering delay $\tau$ and of its
estimate $\tilde{\tau}$ for $N=4$ using $\alpha=0.2$. }
\label{fig:test_delay-1}
\end{figure}

In parallel, the parameter $\alpha$ in~\eqref{eq:AvgRateEstim} is
tuned to provide the best estimate of the buffering delay. Figure~\ref{fig:alpha}
represents the means square error $\text{MSE}(\tilde{\tau},\tau)$
between the actual buffering delay $\tau$ and the estimated one $\tilde{\tau}$
as a function of $\alpha$. The value $\alpha=0.2$ provides the best
estimate.
The evolution with time of the actual buffering delay $\tau$ and
of its estimate $\tilde{\tau}$ is represented in Figure~\ref{fig:test_delay-1}
for $N=4$ using $\alpha=0.2$ and the QF controller for the PSNR
faireness. For this choice of $\alpha$, the estimate provided by
\eqref{eq:AvgRateEstim} for the four video sequences is quite good
for most of the time.

%Robustness of these choices to variations of the PSNR-rate characteristics
%is also verified. As in Section~\ref{Ssec:StabPSNRBufferLevel},
%the location of the roots of $\det\left(z\mathbf{I}-\mathbf{{A}}^{\tau}\right)$
%when the parameter vector $\boldsymbol{a}$ is chosen randomly is
%studied. The same $K=10$ realizations of $N$ random parameter vectors
%of the PSNR-rate model obtained in Section~\ref{Ssec:StabPSNRBufferLevel}
%are considered. The position of the roots corresponding to the video
%characteristic represented in Figure~\ref{fig:video} are provided
%in Figure~\ref{fig:roots1-1}.
%\begin{figure}[h]
%\centering \includegraphics[width=0.95\columnwidth]{Rootlocus_6_PSNR_Delay}
%\caption{Location of the roots of $\det\left(z\mathbf{I}-\mathbf{A}^{\text{\ensuremath{\tau}}}\right)=0$
%for the $K=10$ realizations of the $N=6$ random PSNR-rate characteristics}
%\label{fig:roots1-1}
%\end{figure}
%
%From Figure~\ref{fig:roots1-1}, one sees that all roots remain within
%the unit circle. This result shows that for the choice of the controller
%parameters, the controlled system is reasonably robust to changes
%in the characteristics of the transmitted programs. Here also, some
%of the roots are located quite near the stability limit.

\subsubsection{Results}

The performance of the QF controller using $N=2$, $N=4$, and $N=6$
programs is evaluated in all cases with a transmission rate $R^{\text{c}}=4$~Mps.

We evaluate the PSNR discrepancy $\Delta_{P}$ (in dB) and the PSNR
variance $\sigma_{P}^{2}$ (in dB$^{2}$) defined in~\eqref{eq:sys_perf}.
Additional performance measures are the average delay discrepancy
\begin{equation}
\Delta_{\tau}=\frac{1}{N}\sum_{n=1}^{N}\left(\frac{1}{M}\sum_{l=1}^{M}\left(\tau_{n}(l)-\tau_{0}\right)\right)\label{eq:-3}
\end{equation}
of the buffering delay with respect to $\tau_{0}$ and the variance
of the buffering delay
\begin{equation}
\sigma_{\tau}^{2}=\frac{1}{N}\sum_{n=1}^{N}\left(\frac{1}{M}\sum_{l=1}^{M}\left(\tau_{n}(l)-\tau_{0}-\Delta_{\tau}\right)^{2}\right),\label{eq:-4}
\end{equation}
where $M$ is the number of GoPs in the video streams. Results with
$N=6$ are summarized in Table~\ref{tab:control_delay_video} in
the two cases: QF and TRF controllers.
\begin{table}[htpb]
\centering
\begin{tabular}{|c|c|c|c|c|c|c|}
\hline
 & $K_{\text{P}}^{\text{e,}\tau},K_{\text{I}}^{\text{e,}\tau}$  & $K_{\text{P}}^{\text{t}},K_{\text{I}}^{\text{t}}$ & $\Delta_{\tau}$  & $\sigma_{\tau}^{2}$  & $\Delta_{P}$  & $\sigma_{P}^{2}$ \tabularnewline
\hline
TRF  & $66\times10^3,0$  & $0,0$  & $0.25$  & $0.12$  & $3.8$  & $10.5$ \tabularnewline
\hline
QF  & $66\times10^3,1300$  & $66\times10^3,2600$  & $0.6$  & $0.35$  & $2$  & $10$ \tabularnewline
\hline
\end{tabular}\caption{Performance of QF and TRF controllers when controlling the buffering
delays for $N=6$.}
\label{tab:control_delay_video}
\end{table}
Here again, one notices that
using PI controllers for the transmission rate control loop reduces
the PSNR discrepancy between the programs at the price of some increase
of the buffering delay discrepancy and variance.

\begin{figure}[htpb]
\centering \includegraphics[width=\columnwidth]{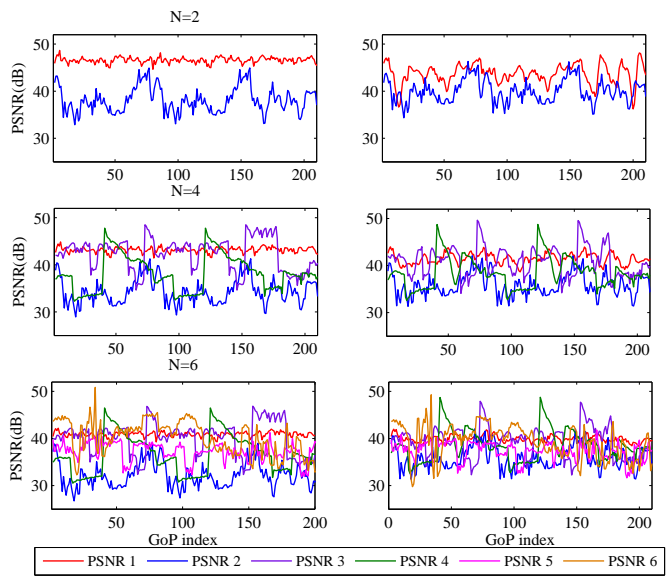}
\caption{Evolution of the PSNR for $N=2$, $N=4$, and $N=6$ using
TRF (left) and QF (right) controllers when controlling the buffer
delays. }
\label{fig:psnr_delay}
\end{figure}

Figure~\ref{fig:psnr_delay} represents the evolution of the PSNR
when considering the TRF controller (left) and the proposed QF controller
(right) $N=2$, $N=4$, and $N=6$ programs. The proposed QF controller
reduces the PSNR discrepancy between the $N$ programs compared to
the TRF controller. Compared to Figure~\ref{fig:psnr_bits}, the
control with the buffer level appears to be less reactive. For example,
when $N=2$, to improve the PSNR of the second program, the PSNR of
the first program has to be decreased. In Figure~\ref{fig:psnr_bits},
the PSNRs are almost immediately adjusted. This is done with some
delay in Figure~\ref{fig:psnr_delay}. This may be due to the difficulty
to accurately estimate the buffering delay. A better response could
be obtained by increasing $K_{\text{P}}^{\text{e,}\tau}$, which relates
the buffering delay and the encoding rate. This, however, would be
at the price of a loss in robustness of the global system to variations
of the PSNR-rate characteristics of the programs.

\begin{figure}[htpb]
\centering \includegraphics[width=\columnwidth]{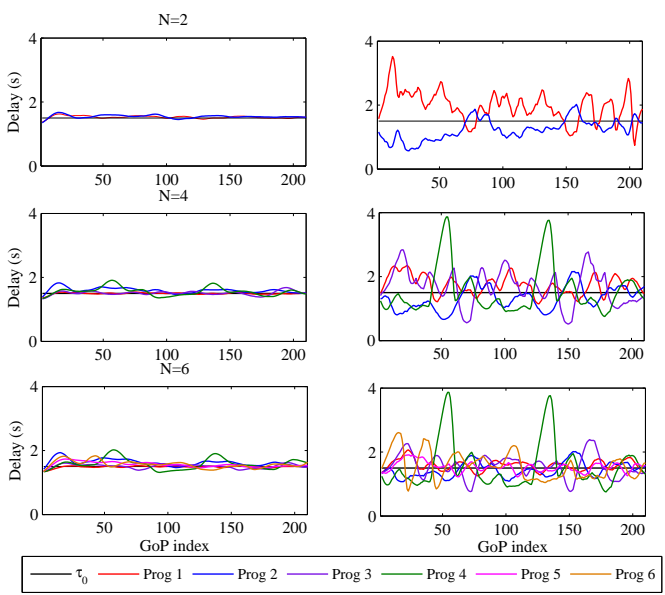}
\caption{Evolution of the buffering delay for $N=2$, $N=4$, and $N=6$ using
TRF (left) and QF (right) controllers when controlling the buffer
delays. }
\label{fig:buffer_bits-1}
\end{figure}

Figure~\ref{fig:buffer_bits-1} represents the evolution of the buffering
delays of $N$ programs of $300$ GoPs using the TRF (left) and the QF
(right) controllers for $N=2$, $N=4$, and $N=6$. With the TRF controller,
the buffering delays reach rapidly $\tau_{0}$ and show a reduced
variance compared to a system with a QF controller. The larger variations
of the buffering delay for the QF controller are due to the interactions
of both control loops (encoding rate and transmission rate). Again,
$K_{\text{P}}^{\text{e,}\tau}$ appears to be too low: large deviations
of the buffering delay are required to reach PSNR fairness.

\subsection{Robustness of the proposed solution to variations of the number of
users and of the channel rate}

In this section, the robustness of the proposed control system (buffering
delay control) is evaluated with respect to variations of the channel
rate and of the number of transmitted video programs. Similar results
are obtained when buffer levels are controlled.

\begin{figure}[t!h]
\centering \includegraphics[width=\columnwidth]{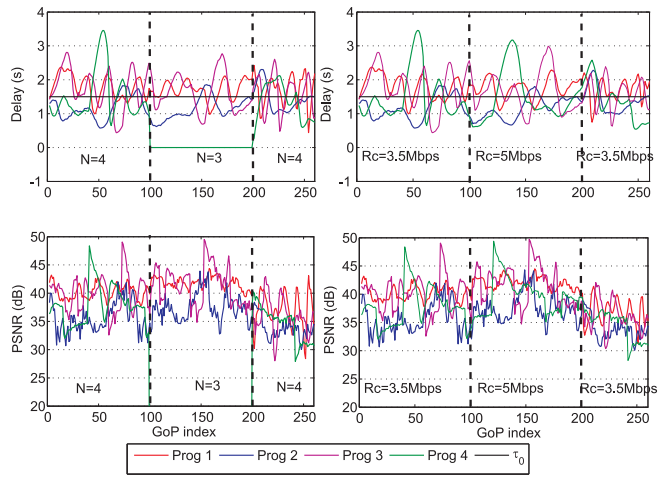}
\caption{System performance using PI controllers while multiplexing four video
programs using the proposed QF controller when considering variations
of the channel rate (left) and of the number of programs (right).}
\label{fig:test_delay}
\end{figure}

First, the number $N$ of transmitted video programs evolves with
time (left). Second, the rate of the channel switches between $R^{c}=3.5$~Mbits/s
and $R^{c}=5$~Mbits/s (right), see Figure~\ref{fig:test_delay}.
The PSNR is used as quality measure. When a new video program is transmitted,
initially, it has no transmission rate allocated by the MANE (since
at time $j$ the controller derives the encoding rate for time $j+1$).
Thus, we choose to set the encoding rate at that time as $R^{\text{c}}/N$.
In Figure~\ref{fig:test_delay} (left), Prog~$4$ is not transmitted
between GoP $35$ and $65$. The same values for the gains of the
PI controllers are used here as in Section~\ref{Ssec:BufferingDelayControlEx}.

%Since the stability of the system depends on the channel rate $R^{c}$ and on the number of video programs $N$, the PI parameters are set so that the system is stable for $N=3$ and $N=4$ and for $R^{c}=3.5$~Mbits/s and $R^{c}=4.5$~Mbits/s. The parameters of the PI controllers are the same as in the previous section.

When the channel rate increases or when a video program is no more
transmitted, the bandwidth allocation adapts rapidly to this change
by providing more rate to programs with low video quality (here Prog~$2$).
When the channel rate decreases or when a new video program is transmitted,
the bandwidth allocation performs well, showing the robustness of
the proposed control system to variations of the channel rate and
to the number of transmitted video programs.

\section{Conclusions and perspectives}
\label{sec:conclu}

In this paper, we propose encoding and transmission rate controllers
for the transmission of several video streams targeting similar video
quality between streams as well as efficient control of the buffering
delay. The controlled system is modeled with a discrete-time non-linear
state-space representation. PI controllers for the transmission rate
and the encoding rate control are considered. The delay introduced
by the network propagation between the MANE and the encoders is taken
into account. This allows to test the stability of the control system
in presence of feedback delay. Simulation results show that the quality
fairness (measured with PSNR or SSIM) is improved compared to a solution
providing an equal transmission rate allocation. Moreover, the jitter
of the buffering delay remains reasonable. The robustness to variations
of the characteristics of the channel and of the number of transmitted
programs has been shown experimentally.

Simulations are performed at the GoP granularity. Control at the frame
level should be considered, however this may require a better consideration
of the communication delay between the MANE and the encoders which
may be variable with the time. This would also require to better account
for the delay, which significantly impedes the behavior of the global
control system, especially when controlling the buffering delay. Tools
devoted to the control of time-delay systems may be useful in this
context, see, \emph{e.g.},\cite{Michiels2007}.

\begin{IEEEbiography}[{\includegraphics[width=1in,height=1.25in,clip,keepaspectratio]{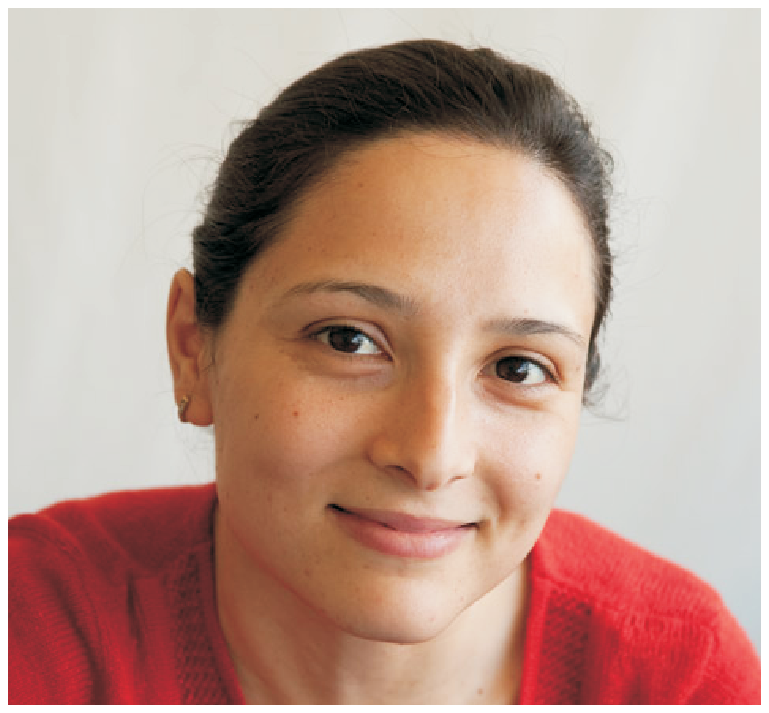}}]{Nesrine Changuel}
(IEEE S'09 - M'12) received her B.S. degree in 2006 and Eng. degree and M.S.
degree in electrical engineering in 2008 from the Ecole Nationale Supérieure d'Electronique et de Radioélectricitée
de Grenoble, France.
She obtained a PhD degree in Control and Signal Processing in 2011 from the Paris-Sud University, Orsay.
She is currently working as a research engineer in Alcatel Lucent Bell Labs. Her current research interests are in the areas of video coding, Statistical multiplexing of video programs, scalable video, Rate and Distortion model, resource allocation and scheduling, Markov decision processes (MDPs), and reinforcement learning. She is a Member of the IEEE.

\end{IEEEbiography}

\begin{IEEEbiography}[{\includegraphics[width=1in,height=1.25in,clip,keepaspectratio]{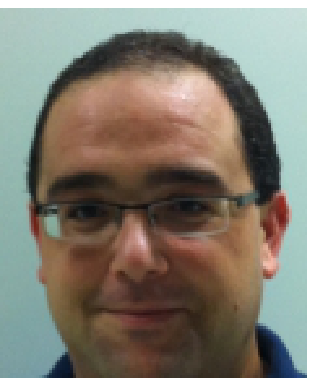}}]{Bessem Sayadi}
is a member of technical staff in Multimedia Technology domain at Alcatel-Lucent Bell Labs, France. He is member of Alcatel-Lucent Technical Academy, since 2008. He received M.Sc. (00) and Ph.D. (03) degrees in Control and Signal processing from Sup\'elec, Paris-Sud University, with highest distinction. He worked previously as a postdoctoral fellow in the National Centre for Scientific Research (CNRS), and as a senior researcher engineer in Orange Labs. His main research interests are in the area of broadcast technology (DVB, 3GPP), video coding and transport, and resource allocation algorithms for communication networks. He has authored over 55 publications in journal and conference proceedings and serves as a regular reviewer for several technical journals and conferences. He holds nine patents and has more than twenty patent applications pending in the area of video coding and wireless communications.
\end{IEEEbiography}

\begin{IEEEbiography}[{\includegraphics[width=1in,height=1.25in,clip,keepaspectratio]{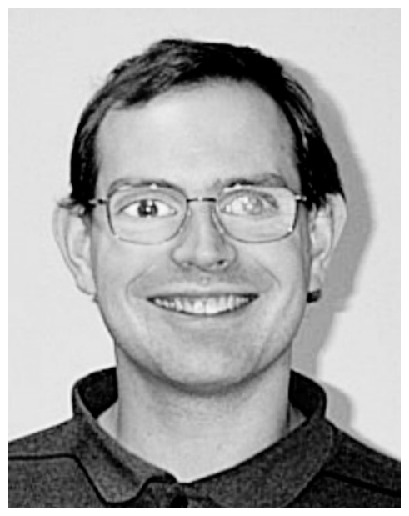}}]{Michel Kieffer}
(IEEE M'02 - SM'07) received in 1995 the Agrégation in Applied Physics at the Ecole Normale Supérieure de Cachan. He obtained a PhD degree in Control and Signal Processing in 1999, and the Habilitation à Diriger des Recherches degree in 2005, both from the Paris-Sud University, Orsay.

Michel Kieffer is a full professor in signal processing for communications at the Paris-Sud University and a researcher at the Laboratoire des Signaux et Systèmes, Gif-sur-Yvette. Since 2009, he is also invited professor at the Laboratoire Traitement et Communication de l'Information, Télécom ParisTech, Paris.

His research interests are in signal processing for multimedia, communications, and networking, distributed source coding, network coding, joint source-channel coding and decoding techniques, joint source-network coding. Applications are mainly in the reliable delivery of multimedia contents over wireless channels. He is also interested in guaranteed and robust parameter and state bounding for systems described by non-linear models in a bounded-error context.

Michel Kieffer is co-author of more than 120 contributions in journals, conference proceedings, or books. He is one of the co-author of the book Applied Interval Analysis published by Springer-Verlag in 2001 and of the book Joint source-channel decoding: A crosslayer perspective with applications in video broadcasting published by Academic Press in 2009.
He is associate editor of Signal Processing since 2008, of the Journal of Communication and Information Systems since 2011 and of the IEEE Transactions on Communications since 2012.
In 2011, Michel Kieffer became junior member of the Institut Universitaire de France.

\end{IEEEbiography}

\end{document}